\documentclass[12pt]{article}
\pdfoutput=1
\usepackage[utf8x]{inputenc}
\usepackage[ inner=1cm, outer=1cm, top=1cm, bottom=1.9cm]{geometry}
\usepackage[svgnames]{xcolor}
\usepackage[unicode]{hyperref}
\usepackage[bottom]{footmisc}
\usepackage{graphicx}
\usepackage{framed}
\usepackage{csquotes}
\usepackage{tikz}
\usetikzlibrary{decorations.markings}
\usetikzlibrary{decorations.pathmorphing}
\definecolor{shadecolor}{rgb}{0.90,0.90,0.90}
\usepackage{hyperref}
\usepackage{subcaption}
\usepackage{pifont}
\usepackage{setspace}
\usepackage{amsmath, amssymb, amsthm, float, graphicx,amsfonts}
\numberwithin{equation}{section}

\hypersetup{colorlinks=true, linkcolor=Maroon, citecolor=Maroon,urlcolor=IndianRed,linktocpage}
\def\beq{\begin{eqnarray}}\def\eeq{\end{eqnarray}}
\def\be{\begin{equation}}\def\ee{\end{equation}}
\def\p{\pi}
\def\g{\gamma}

\def\m{\mu}
\def\n{\nu}
\def\a{\alpha}

\def\b{\beta}
\def\d{\delta}
\def\c{\chi}
\def\f{\phi}
\def\t{\theta}
\def\D{\Delta}
\def\F{\Phi}
\def\G{\Gamma}
\def\l{\lambda}
\def\L{\Lambda}
\def\x{\xi}
\def\pd{\partial}

\def\bz{\bar{z}}

\def\mm{{\mathcal{M}}}
\def\mi{{\mathcal{I}}}

\def\ml{{\mathcal{L}}}

\def\G{\Gamma}

\def\w{\omega}

\def\Tr{{\rm Tr}}

\usepackage{bm}
\usepackage{bm}

\begin{document}

\title{\bf{ On the Regge limit of Fishnet correlators}}
\author{Subham Dutta Chowdhury${}^\nu$\footnote{subham@theory.tifr.res.in}, ~Parthiv Haldar${}^J$\footnote{parthivh@iisc.ac.in}~ and~ Kallol Sen ${}^h$\footnote{kallolmax@gmail.com}\\ ~~~~\\
\it	${}^{\nu}$Tata Institute of Fundamental Research\\ 
\it	Homi Bhabha Road, Navy Nagar, Colaba, Mumbai 400005, India,\\ 
\it	${}^{J}$Center for High Energy Physics,Indian Institute of Science\\
\it	C.V. Raman Road, Bangalore 560012, India,\\
\it	${}^{h}$Kavli Institute for the Physics and Mathematics of the \it Universe (WPI),\\
\it	University of Tokyo, Kashiwa, Chiba 277-8583, Japan.}

\maketitle

\abstract{We study the Regge trajectories of the Mellin amplitudes of the $0-,1-$ and $2-$ magnon correlators of the Fishnet theory. Since fishnet theory is both integrable and conformal, the correlation functions are known exactly. We find that while for $0$ and $1$ magnon correlators, the Regge poles can be exactly determined as a function of coupling, $2$-magnon correlators can only be dealt with perturbatively. We evaluate the resulting Mellin amplitudes at weak coupling, while for strong coupling we do an order of magnitude calculation.}

\tableofcontents

\onehalfspacing

	

\section{Introduction}
\paragraph{} 

 $\mathcal{N}=4$ SYM is one of the few most convenient playground for analyzing the scattering amplitudes for a CFT, since in addition to conformal symmetries, it also admits a Lagrangian description. But this has its own technical challenges. A somewhat simpler theory is obtained from the $\g-$deformed $\mathcal{N}=4$ SYM in the double scaling limit, called the conformal fishnet theory. In this limit, all the heavier constituents of the $\mathcal{N}=4$ except the adjoint scalars decouple (their interaction with the retained scalars is tuned to zero), giving an effective Lagrangian \cite{Gurdogan:2015csr},
\be\label{lagfish}
\mathcal{L}= N~ \Tr [\pd_\mu\bar{X}\pd^\m X+\pd_\mu\bar{Z}\pd^\m Z+(4\pi \xi)^2 X Z \bar{X}\bar{Z}]\,,
\ee
where $X(Z)$ are complex traceless $N\times N$ adjoint scalars and $\bar{X}(\bar{Z})$ are the conjugates. The reduced coupling $\xi$ is given in the planar limit ($N\rightarrow\infty$, $g_{YM}^2\rightarrow0$) for specific configuration of the deformation ($\g_3\rightarrow i\infty$) by \cite{Gurdogan:2015csr}, 
\be
\xi^2= g_{YM}^2 N e^{-i\g_3}=\text{finite}\,.
\ee
Due to CPT non-invariance of the interaction term, the theory is inherently non-unitary giving rise to some peculiar features. Owing to integrability and conformal invariance, correlation functions of local and bi-local operators in this theory can be exactly determined as a function of the coupling $\xi$, by iteratively solving the Bethe-Salpeter equations (reviewed in detail in section \ref{fishnet}). Authors of  \cite{Gromov:2018hut,Korchemsky2018} analyzed the scattering amplitudes for the fishnet theory in four dimensions. Further, one can analyze the Regge limit of the correlators of the fishnet theory exactly in coupling $\xi$\footnote {Unlike other theories, where the Regge trajectories are only known in certain limits (say the weak coupling limit), here the trajectories are exact functions of the coupling $\xi$.}. The exact correlation functions of the local and bi-local operators that we study is given by (from \eqref{correlationtomagnon}),
\begin{eqnarray}
\langle\Tr\left(X(x_1)X(x_2)\right)\Tr\left(\bar{X}(x_3)\bar{X}(x_4)\right)\rangle &=& G^{0}(x_1,x_2|x_3x_4)+G^{0}(x_1,x_2|x_4x_3),\nonumber\\
\langle\Tr\left(X(x_1)Z(x_1)X(x_2)\right)\Tr\left(\bar{X}(x_3)\bar{X}(x_4)\bar{Z}(x_4)\right)\rangle &=& \frac{1}{2}G^{1}(x_1,x_2|x_3x_4)-\left(\xi^2 \rightarrow -\xi^2 \right),\nonumber\\
\langle\left({\cal O}_{XZ}(x_1){\cal O}_{XZ}(x_2){\cal O}_{\bar{X}\bar{Z}}(x_3){\cal O}_{\bar{X}\bar{Z}}(x_4)\right)\rangle &=& G^{2}(x_1,x_2|x_3x_4),
\end{eqnarray}
where ${\cal O}_{XY}(z)=\text{tr}~(XY)(z)$. These correlation functions are expressed in terms of the $n$-magnon graphs denoted by $G^{n}(x_1,x_2|x_3x_4)$.

We are interested in studying the Regge limit of the correlators appearing in this theory. The Regge limit for a scattering process in a theory is defined as a special kinematic limit of $2 \rightarrow 2$ scattering of particles in which the Centre of Mass (COM) momenta is taken to be large. In terms of mandelstam variables $s ,t$ and $u$, this corresponds to large $s$ at fixed $t$. Regge scattering has important theoretical and phenomenological aspects for which it serves as an important physical quantity to study \cite{Regge:1959mz}. In particular, the Regge limit of the scattering amplitude encodes information about the spectrum of the exchanged particles. The leading Regge trajectory is governed by the particle with the highest spin that is being exchanged (also referred to as Reggeon) and hence does not require full knowledge of the spectrum. 
\be
A_{\text{Regge}}(s,t)\sim s^J\,.
\ee
There are several interesting examples where such studies have been undertaken, In the context of String theory, the Virasoro-Shapiro amplitude, which describes the scattering amplitude for 4 dilatons in type II Superstring theory \cite{PhysRev.177.2309}, the Regge limit of the scattering amplitude scales as $$s^{2 + \frac{\a' t}{2}}$$ 
which denotes graviton dominance in the high energy limit (t is negative). Similarly for QCD, one can see from \cite{Korchemsky:1994um} that the LLA (leading log approximation) contribution to the Regge limit comes from,
\be
J=1+\D_{\text{BFKL}}(t)\,,
\ee
The same can be shown in a perturbative manner for the $\mathcal{N}=4$ SYM \cite{Costa2012a} for which in weak coupling, 
\be
J=2-\text{subleading}\,.
\ee
In contrast, for the fishnet theories under consideration, we find that for the $0,1,2-$magnon cases, in the weak coupling, the leading Regge theory is dominated by,
\be
J=0,-1,-2\,,
\ee
respectively. This is expected to be connected with the inherent non-unitarity of the theory so that the effective exchanges in the Regge limit has negative spins. In this case, the LLA contribution is expected to come effectively from the $0-$magnon graphs, in a simple form,
\be
A_{LLA}(s,t)\propto \log s\,,
\ee

In \cite{Korchemsky2018}, the author studies the Regge limit of the $0-$magnon four point amplitude in the fishnet theory using standard LSZ reduction techniques in momentum space. 
 An immediate obstruction to generalizing their method to the $1$ and $2$- magnon cases is the fact that the $1$ and $2$- magnon states describe a bound state which is off-shell. What is meant by this is that,  the external operators for $1$ and $2$-magnon cases can not be put on-shell. For example, $1$-magnon state $XZ(x)$ after a Fourier transform describes a two-particle state that cannot be on-shell. Another way to see this is to verify that $4$-point correlator $\langle\Tr[XZ(x_1)X(x_2)]\Tr[\bar{X}\bar{Z}(x_3)\bar{X}(x_4)]\rangle$ in the momentum space does not have a pole at $p_1^2=0$ and $p_3^2=0$ (but it does have poles at $p_2^2=0$ and $p_4^2=0$) and, therefore, the LSZ reduction gives a vanishing result. So the technique used in \cite{Korchemsky2018} cannot be used to get Regge amplitudes for the 1 and 2 magnon cases\footnote{We thank Gregory Korchemsky for pointing this out to us.}.

We will however discuss the Regge limit of magnon correlators independently following \cite{Costa2012a}. In \cite{Costa2012a} the authors showed that for the Mellin amplitude for a CFT correlator, given by \cite{Costa2012a},
\be
M(s,t)=\int_{-\infty}^{\infty}d\n \oint \frac{dJ}{\sin \pi J} \ b_J(\n^2) \w_{\n,J}(s,t)P_{\n,J}(s,t)\,,
\ee
where,
\begin{align}
\begin{split}
\w_{\n,J}(s,t)=&\frac{\G(\frac{\D_1+\D_2+J+i\n-h}{2})\G(\frac{\D_3+\D_4+J+i\n-h}{2})\G(\frac{\D_1+\D_2+J-i\n-h}{2})\G(\frac{\D_3+\D_4+J-i\n-h}{2})}{8\pi \G(i\n)\G(-i\n)}\\
&\times \frac{\G(\frac{h+i\n-J-t}{2})\G(\frac{h-i\n-J-t}{2})}{\G(\frac{\D_1+\D_2-t}{2})\G(\frac{\D_3+\D_4-t}{2})}\,,
\end{split}
\end{align}
and $P_{\n,J}(s,t)$ is the Mack polynomial, the Regge limit is defined as $s\rightarrow \infty$ and $t=\text{fixed}$. The details of how the Regge limit is obtained will be discussed in the next section. The most important part is basically the spectral weight $b_J(\n^2)$ which for fishnet CFT can be exactly determined as shown in\cite{Kazakov:2018hrh}. In this short note, we achieve a modest goal of determining the Regge trajectories for the $0,1,2-$magnon correlators in the fishnet CFT using the techniques of \cite{Costa2012a}. We will also point out various relevant features and subtleties of the computations pertaining to each type of correlators. We now present the main results of our paper.

\subsection*{Results}
Our main results can be summarized as follows. We systematically study the Mellin amplitude in the limit of $s \rightarrow \infty$ with $t$ held fixed. $s$ and $t$ are the Mellin variables which are used for $4$-point conformal correlator. These are defined in \eqref{mellvardef} (these are not to be confused with the usual Mandelstam invariants used for $2\to2$ flat space scattering.). For the correlation functions of certain operators in the fishnet theory, We obtain the Regge poles and evaluate the $\nu$ integral in the weak and strong coupling limit for these poles. This is done by considering the Mellin amplitude in the principal series representation as in \eqref{mell}. Using Sommerfeld Watson transform as usually done in studying the Regge limit of QFT scattering amplitudes, we obtain the Regge poles of our correlator. This is presented in detail in section \eqref{crt}.  
\subsubsection*{$0$-Magnon correlator}

The Regge trajectories were evaluated in \cite{Gromov2018} and are given by (as worked out in \eqref{Regge0}),
\begin{align}
\begin{split}
J^\pm_2(\n)=-1+\sqrt{1-\n^2\pm 2\sqrt{f^4-\n^2}}\,,\\
J^\pm_4(\n)=-1-\sqrt{1-\n^2\pm 2\sqrt{f^4-\n^2}}\,,
\end{split}
\end{align}
where, $f=4\sqrt{2}c\p^2\x$. $J(\nu)$ denotes the spin of the Regge pole which we get by deforming the Sommerfeld Watson contour. We have worked out the Mellin amplitude in the Regge limit for weak coupling, $f\rightarrow0$, and strong coupling, $f\rightarrow\infty$ for the leading Regge trajectory $J_2^+(\n)$.\\\\
{\it \underline{Weak coupling:} }
The Mellin amplitude in the Regge limit after the $\nu$ integral is given by,
\begin{align}
\begin{split} 
\mm_{(0)}^\pm(s,t)=\bigg[\pm2c^4f^2 \frac{1}{4} &(q (\pi  \pmb{L}_1(q)+2) I_0(q)-(\pi  q \pmb{L}_0(q)+2) I_1(q))+\ldots\bigg]\pm\,(s\rightarrow -s)\,,
\end{split}
\end{align}
where $q=f^2\log(s/4)$ and $I_n(q),~ \pmb{L}_m(q)$ are respectively Modified Bessel function of first kind and Modified Struve function.  The ellipses denote subleading terms. Note that leading terms are independent of $t$. The subleading terms (see \eqref{0magmelReggeweakfinal}) are however $t$-dependent. The limit considered is 
$$f\rightarrow 0,\qquad s\rightarrow \infty, \qquad q=f^2 \log \left(\frac{s}{4}\right) \rightarrow \text{fixed}.$$
{\it \underline{Strong coupling:} } 
	\begin{align}
\begin{split}
\mm_{(0)}^{\pm}(s,t)\sim \left[\mp 2\sqrt[4]{8}c^4\sqrt{\frac{f}{\p}}\csc(\sqrt{2}\p f) \,\frac{s^{\sqrt{2}f}}{s\log ^{\frac{3}{2}}(s)}\frac{\G\left(\frac{3-t-\sqrt{2}f}{2}\right)^2}{\G\left(1-\frac{t}{2}\right)^2}\right]+\,(s\rightarrow -s).
\end{split}
\end{align}
\subsubsection*{$1$-Magnon correlator}
For this case there are two separate Regge trajectories depending upon whether it is even or odd spin. We have used the following definitions below $$q=\log(s),\qquad g=8\p^2 c\x.$$
\textit{\underline{Even Spin:}} 
The Regge trajectory is,
\be 
J^{\pm}_{e}=-1\pm\sqrt{g^2-\n^2}.
\ee 
The Mellin amplitudes for strong coupling and weak coupling are as following.
\begin{itemize}\item{\it Weak coupling:}\\
	\begin{align}\label{wce}
	\begin{split}
	&\mm^{+}_{(1)}(s,t)= -\frac{8c^4g^2}{s}
	\bigg[
	\frac{  I_1(q)}{q}
	-g\frac{  I_2(q)}{q}\left\lbrace\psi ^{(0)}\left(\frac{3}{2}-\frac{t}{2}\right)+\log (4)\right\rbrace+O(g^2)\bigg]+(s\rightarrow-s),
	\end{split}
	\end{align}
	where we have considered the limit,
	$$s \rightarrow \infty, \qquad g\rightarrow 0,\qquad q=g\log s \rightarrow \text{constant}.$$
	Further $I_n(x)$ is Modified Bessel function of first kind. Note that here also, the leading term is $t$-independent.
	\item{\it Strong coupling:}
	\begin{align}\label{sce}
	\begin{split}
	\mm^+_{(1)}(s,t)\sim \left[-\frac{4\sqrt{2}c^4}{\sin(\p g)}\sqrt{\frac{g}{\pi }} \, \frac{s^{g} }{ s\log ^{\frac{3}{2}}(s)}\frac{\G\left(\frac{3-t-g}{2}\right)^2}{\G\left(\frac{3-t}{2}\right)^2}\right]+(s\rightarrow-s). 
	\end{split}
	\end{align} 
\end{itemize}
\textit{\underline{Odd Spin:}} 
The Regge trajectory is given by,
\be 
J^{\pm}_o=-1\pm i\sqrt{g^2+\n^2}
\ee while the Mellin amplitudes are, 
\begin{itemize}\item {\it Weak coupling:}
	\begin{align}\label{wco}
	\begin{split}
	\mm_{(1)}^-=&-\frac{4c^4}{\pi s}\left[g^2\frac{ \pi  J_1(q)}{q}+ g^3   \left\lbrace\psi ^{(0)}\left(\frac{3}{2}-\frac{t}{2}\right)+\log (4)\right\rbrace\frac{\pi  J_2(q)}{q}\right]+O(g^4)-(s\rightarrow-s).
	\end{split}
	\end{align}
	where we have considered the following limit,
	$$s \rightarrow \infty, \qquad g\rightarrow 0,\qquad q=g\log s \rightarrow \text{constant}.$$ and $J_m(x)$ is Bessel function of first kind. 
	\item {\it Strong coupling:}
	\begin{align}\label{sco}
	\begin{split} 
	\mm_{(1)}^-\sim \left[\frac{4 c^4 (1+i)}{s\log^{\frac{3}{2}}(s)}\text{csch}(\pi  g)\sqrt{\frac{g}{\p}}\left\lbrace is^{ig}\,\frac{\G\left(\frac{3-t-ig}{2}\right)^2}{\G\left(\frac{3-t}{2}\right)^2}-s^{-ig}\,\frac{\G\left(\frac{3-t+ig}{2}\right)^2}{\G\left(\frac{3-t}{2}\right)^2}\right\rbrace\right]-(s\rightarrow -s).
	\end{split} 
	\end{align} 
\end{itemize}
\subsubsection*{$2$-Magnon Correlator}
For the 2-magnon case we have evaluated the Regge trajectories as well as the Mellin amplitudes perturbatively in weak coupling and strong coupling limits. The main results for this case are as following,\\
{\itshape \underline{Weak coupling:}} The leading Regge trajectory in this case is given by,
\be
J(\n)=\begin{cases} -2+i\n-\sum_{k\geq1}\xi^{2k}\g_{0,k}\,, |\n|>\xi^4;\\-2+ \alpha _1 \xi ^{4/3}+\frac{1}{3} \alpha _1^2 \xi ^{8/3}+\frac{1}{3} \alpha _1^3 \xi ^4+\frac{\alpha _1^4 \xi ^{16/3} (1120-81 \zeta (3))}{2592}+\cdots\end{cases}
\ee 
with $\g_{0,k}$ being given explicitly in \eqref{2magReggeweak}. The Mellin amplitude in this case is given by,
\begin{align}
\begin{split}
\mm_{(2)}^+&=L^{-2}
\left\lbrace
\left[
-\xi ^{32/3}\frac{1}{2304 \pi ^{9}}
+\xi ^{12}\frac{17 }{576 \pi ^{9}}\psi ^{(0)}\left(2-\frac{t}{2}\right)
+\frac{\x^{40/3}}{5184\p^9}
\left(12\psi ^{(0)}\left(2-\frac{t}{2}\right)-18\psi ^{(0)}\left(2-\frac{t}{2}\right)^2
\right.
\right.
\right.\\
&\left.
\left.
\hspace{2.3 cm}-9\psi ^{(1)}\left(2-\frac{t}{2}\right)-9\p^2-8\right)+\dots
\right]
+\log (L)\left[-\frac{\xi ^{12} }{288 \pi ^{9}}+\frac{\x^{40/3}}{216} +\dots \right]
\\
&\left.\hspace{7.8 cm}+\log^2L\left[-\frac{\x^{40/3}}{288\p^9}+\dots\right]+O(\log^3L)
\right\rbrace+~(s\to-s)
\end{split}
\end{align}
where $L=\frac{s}{4}$. $\mm_{(2)}^-$ is zero for $2$-magnon because the amplitude is symmetric under $s \rightarrow -s$. 
\\
{\itshape \underline{Strong coupling :}} In strong coupling the leading Regge trajectory is given by,
\begin{align}
\begin{split}
J&=-1+\left[2\sqrt[4]{2}\xi -\frac{\nu ^2+3}{4 \sqrt[4]{2}\xi}+\frac{87+18 \nu ^2-\nu ^4}{64 \sqrt[4]{8}\xi^3}+\mathcal{O}\left(\frac{1}{\xi^4}\right)\right]
\end{split}
\end{align}
and the corresponding Mellin amplitude is given by,
\begin{align}
\begin{split}
M_{(2)}^{+}&\sim \left[-\frac{1}{16\ 2^{7/8}\p^{10}}\sqrt{\frac{\x}{\p}}\frac{s^{2 \sqrt[4]{2} \xi  }}{s\log^{\frac{3}{2}}s}\csc \left(2 \sqrt[4]{2} \pi  \xi \right)\frac{\G\left(\frac{3-t}{2}-\sqrt[4]{2}\x\right)^2}{\G\left(\frac{4-t}{2}\right)^2}\right]+ (s\to-s).
\end{split}
\end{align}
\paragraph{}The paper is organized as follows. In section \ref{fishnet}, we discuss the basics of the fishnet CFT in four dimensions following \cite{Grabner:2017pgm,Gromov:2018hut,Gurdogan:2015csr}. In section \ref{crt}, we give a brief overview of the ``{\it Conformal Regge Theory}" following \cite{Costa2012a,Cornalba:2007fs,Cornalba:2008qf}. Specifically, we elaborate a bit on the pole analysis and the contour prescription associated with the resultant Mellin amplitude in the Regge-limit. In sections \ref{0mag}, \ref{1mag} and \ref{2mag}, we discuss the application of the Conformal Regge theory to the case of the fishnet correlators. We discuss in details the Regge trajectories associated with the individual types of magnon correlators. For $0$ and $1-$magnon, we compute the Mellin amplitudes for the leading Regge trajectories both in the weak and strong coupling regimes. For $2-$magnon case,  we analyze the systematics of the Regge limit separately in the weak and strong coupling regimes. We end the paper with some discussions on what could be the potential issues and further questions. In Appendix \ref{polestructure}, we give the details of the assumptions specially the pole analysis and contour prescription along the lines of \cite{Korchemsky2018} for individual cases. In Appendix \ref{intweak}, we provide the details of the integrals. We demonstrate that there is only one integral per case one needs to compute and the subsequent integrals (for the weak coupling systematics) are just finite integrals with respect to one of the Mellin variables. In Appendix \ref{2magdetails}, we provide a separate discussion of the $2-$magnon case in the weak and strong coupling regime. 

\section{Conformal Fishnet theory in $4d$}\label{fishnet}
\paragraph{}In this section we review the Bi-scalar fishnet CFT \cite{Gurdogan:2015csr} and provide an overview of the basic structure of the correlation functions that can be exactly computed in the planar limit \cite{Gromov:2018hut}. The Bi-scalar fishnet CFT is obtained as the double scaling limit of the $\gamma$ deformed ${\cal N}=4$ Super Yang-Mills \cite{Gurdogan:2015csr}. The $\gamma$-deformation reduces the $SU(4) \sim SO(6)$ ${\cal R}$-symmetry of the theory to $U(1)^3$. The double scaling limit is defined as $\gamma_i \rightarrow \infty, g^2=N_cg_{ym}^2 \rightarrow 0$ with $\zeta_j^2=g^2e^{-i\gamma_j}$ held fixed (where $i=1,2,3 $ are the three cartans of $SO(6)$). Choosing $\zeta_1,\zeta_2 \rightarrow 0$, all the other fields except two complex scalars decouple and we obtain the classical Lagrangian for the Bi-Scalar CFT given by \eqref{lagfish}.  At the quantum level, the theory described by this Lagrangian is not conformal and we need suitable double trace counter terms \cite{Fokken:2013aea,Grabner:2017pgm}. The exact details of these counter terms will not be important for our analysis. The theory with the counter terms is renormalizable and has non-trivial fixed points where the coupling constants of the counter terms can be described as (complex) functions of the coupling constant $\xi$. 
The theory at the fixed point is conformal and integrable in the planar limit \cite{Sieg:2016vap,Zamolodchikov:1980mb,Chicherin:2017cns,Gromov:2017cja}. The resulting theory is non-unitary and conformal. One can consider correlation functions of the local protected dimension 2 and bi-local operators such as 
\be
{\cal O}_{xz}(x) =\text{\Tr}\left( XZ \right) (x), \qquad {\cal O}_{xzx}(x_1,x_2) =\text{\Tr}\left( X(x_1)Z(x_1)X(x_2) \right). 
\ee 
It was shown in \cite{Gromov:2018hut} that due to the iterative structure of the Feynman graphs that contribute to the unprotected four point functions that can be built out of these operators, they can be computed exactly in the planar limit. These correlation functions exhibit a rich non-perturbative OPE structure. We briefly recall the salient features  of their computation. The building blocks for the correlation functions are termed as \enquote{$n$-magnon} correlators, denoted by $G^{n}(x_1,x_2|x_3x_4)$, depending on the particle that is being exchanged.
The relation between the magnon graphs and actual correlation functions are given below \cite{Gromov:2018hut}.
\begin{eqnarray}\label{correlationtomagnon}
\langle\Tr\left(X(x_1)X(x_2)\right)\Tr\left(\bar{X}(x_3)\bar{X}(x_4)\right)\rangle &=& G^{0}(x_1,x_2|x_3x_4)+G^{0}(x_1,x_2|x_4x_3),\nonumber\\
\langle\Tr\left(X(x_1)Z{x_1}X(x_2)\right)\Tr\left(\bar{X}(x_3)\bar{X}(x_4)\bar{Z}(x_4)\right)\rangle &=& \frac{1}{2}G^{1}(x_1,x_2|x_3x_4)-\left(\xi^2 \rightarrow -\xi^2 \right),\nonumber\\
\langle\left({\cal O}_{XZ}(x_1){\cal O}_{XZ}(x_2){\cal O}_{\bar{X}\bar{Z}}(x_3){\cal O}_{\bar{X}\bar{Z}}(x_4)\right)\rangle &=& G^{2}(x_1,x_2|x_3x_4)
\end{eqnarray}   
\paragraph{} The 0-1 and 2 magnon graphs have the periodic "fishnet" structure and can be computed using the Bethe-Salpater approach. In terms of the iterative Feynman diagram structure, they can be written down as \cite{Gromov:2018hut} \footnote{The periodic structure as well as the nomenclature is evident from the pictorial representation of these correlators presented in figure 1 and figure 5 of \cite{Kazakov:2018hrh}},
\begin{eqnarray}
G^{0}(x_1,x_2|x_3x_4) &=& \sum_{n\geq 0} (16\pi^2\xi^2)^n G^{0}_n(x_1,x_2|x_3x_4),\nonumber\\
G^{1}(x_1,x_2|x_3x_4) &=& \sum_{n\geq 0} (16\pi^2\xi^2)^n G^{1}_n(x_1,x_2|x_3x_4),\nonumber\\
G^{2}(x_1,x_2|x_3x_4) &=& \sum_{n\geq 0} (16\pi^2\xi^2)^2n G^{2}_n(x_1,x_2|x_3x_4).\nonumber\\
\end{eqnarray} 
 The actual procedure for evaluating these summed diagrams involves expressing these in terms of a graph building operator $\hat{H}$. Schematically, the correlator 

\be\label{graphgen}
(x_1,x_2|x_3x_4) \sim \langle x_1,x_2| \hat{G}|x_3,x_4\rangle,\qquad \hat{G}\sim\sum_{i=0}^\infty f(\xi)^i \hat{H}^{n+i}. 
\ee  More precisely, since $\hat{H}$ commutes with the conformal group, the eigenstate $\langle x_1,x_2|$ is basically the three point functions of two scalar operators of dimension $\Delta_1 $ and$ \Delta_2$ at position $x_1$ and $x_2$ and some spin $J$ operator with $\Delta=2+ i \nu$ at $x_0$.  
The eigenvalue equation satisfied by $\hat{H}$ is then given by,
\begin{eqnarray}
\int d^d x_1 d^d x_2 \hat{H}(x_1,x_2,x_3,x_4)\Phi_{J,\nu,x_0}(x_1,x_2)=E_{\D,J}\Phi_{J,\n,x_0}(x_3,x_4),
\end{eqnarray}
where $E_{\D,J}$ are the eigenvalues of the graph building operator. These eigenfunctions are the conformally invariant three point functions,
\be
\Phi_{J,\nu,x_0}(x_1,x_2)=\frac{2^J}{x_{12}^{\D_1+\D_2-\D+J}x_{10}^{\D_{12}+\D-J}x_{20}^{\D-J-\D_{12}}}\bigg(\frac{n\cdot x_{02}}{x_{02}^2}-\frac{n\cdot x_{01}}{x_{01}^2}\bigg)^J\,,
\ee
projected onto a light-like (null) vector $n_{\mu}$. We can then write the graph-building operator as,
\be
\hat{H}(x_1,x_2,x_3,x_4)=\sum_{J=0}^\infty\frac{(-1)^J}{(x_{12}^2)^{\D_1+\D_2-4}}\int_0^\infty \frac{d\n}{c_1(\n,J)}E_{\D,J}\int d^4x_0 \Phi^{\m_1\dots\m_J}_{-\n,x_0}(x_1,x_2)\Phi^{\m_1\dots\m_J}_{\n,x_0}(x_3,x_4)\,,
\ee
where the function $c_1(\n,J)$ in arbitrary dimensions is given by \cite{Dobrev:1977qv},
$$c_1(\n,J)= \frac{2^{J+1}J! \Gamma(i\nu)\Gamma(-i\nu)(\nu^2 + (\frac{d}{2}+J-1)^2)^{-1}}{\pi^{-\frac{3d}{2}+1}\Gamma(\frac{d}{2}-1+i\nu)\Gamma(\frac{d}{2}-1-i\nu)\Gamma(\frac{d}{2}+J)}.$$
The last integral can be put in terms of the familiar conformal block and its shadow {\it viz} \cite{Dolan:2000ut, Dolan:2003hv, Dolan:2011dv}, 
and finally from \eqref{graphgen}, \cite{Gromov:2018hut}
\be\label{4pt}
G(x_1,x_2,x_3,x_4)=\sum_{J=0}^\infty\frac{(-1)^J}{(x_{12}^2)^{\D_1+\D_2-4}}\int_0^\infty \frac{d\n}{c_2(\n,J)}\frac{\left(E^{(n)}_{\n,J}\right)^p}{1-\chi_n E^{(n)}_{\n,J}}\ g_{\n,J}(z,\bz)\,,
\ee
where $p=1,2,1$ for $n=0,1,2-$magnon graphs respectively and $c_2(\n,J)=c_1(\n,J)/c(\n,J)$ and is given by \cite{Dobrev:1977qv},\\
\begin{align} 
\begin{split}
c_2(\n,J)&= \frac{
	2\pi^{d+1}J! 
	\Gamma\left(\Delta-\frac{d}{2}\right)
	\Gamma(\Delta+J-1)
	\Gamma\left(\frac{\d-\Delta+\Delta_1-\Delta_2+J}{2}\right)
	\Gamma\left(\frac{\d-\Delta-\Delta_1+\Delta_2+J}{2}\right)
}
{
	(d-\Delta+J)
	\Gamma(\Delta-1)
	\Gamma\left(\frac{d}{2}+J\right)
	\Gamma\left(\frac{\Delta+\Delta_1-\Delta_2+J}{2}\right)
	\Gamma\left(\frac{\Delta-\Delta_1+\Delta_2+J}{2}\right)
}.
\end{split}
\end{align}
\paragraph{} This is the starting point of our analysis. For more details about the derivation we refer the reader to \cite{Gromov:2018hut}.
Before going into the characterization of the Regge limit for the individual graphs, we will write down the eigenvalues for the $n-$magnon graphs. 
\begin{shaded}
\begin{align}\label{eigs}
\begin{split}
E^{(0)}_{\D,J}&=\frac{16\pi^4c^4}{(J+\D)(J+\D-2)(J-\D+2)(J-\D+4)}\,,\qquad \hspace{1.1 cm}\chi_0=(16\pi^2\xi^2)^2;\\
E^{(1)}_{\D,J}&=(-1)^J\frac{4\pi^2 c^2}{(J+\D-1)(J-\D+3)}\,,\qquad \qquad \hspace{2.7 cm}\chi_1=(16\pi^2\xi^2);\\
E^{(2)}_{\D,J}&=\frac{\psi_1(\frac{J-\D+4}{4})-\psi_1(\frac{J-\D+6}{4})-\psi_1(\frac{J+\D}{4})+\psi_1(\frac{J+\D+2}{4})}{(4\pi)^4(\D-2)(J+1)}\,,\qquad \hspace{0.1 cm}\chi_2=(16\pi^2\xi^2)^2.
\end{split}
\end{align}
\end{shaded}
where $\psi_m(x)=d^m\psi(x)/dx^m$ and $\psi(x)$ is Digamma function given by $\frac{d}{dx}(\ln \G(x))$. In this notation $\psi_0(x)=\psi(x)$.

\section{Conformal Regge theory}\label{crt}
Regge theory is used to describe high energy limit of physical scattering processes. Given a four particle scattering process with Mandelstam invariants $\{S,T,U\}$, 
 $$(p_1+p_2)^2=-S,\qquad (p_1+p_3)^2=-T,\qquad (p_1+p_4)^2=-U,$$ Regge limit correspond to the kinematic regime of large $S$ at fixed $T$. In Regge limit, the leading part of the amplitude is dominated by Regge poles which are functions of actual physical poles of the amplitude. In \cite{Costa2012a} the authors explore an analogy between certain kinematic configurations of conformal correlation functions and Regge limits of flat space scattering amplitudes by studying the correlation functions in the Mellin space. The role of the mandelstam invariants in the scattering is played by the Mellin transform variables $s$ and $t$.
 In this section we review Conformal Regge Theory in Mellin space following \cite{Costa2012a}.  The Mellin representation of a four-point conformal correlator is, 
\be\label{mellvardef}
\mathcal{G}(u,v)=\frac{1}{(4\pi i)^2}\int_{-i\infty}^{i\infty} ds dt\ u^{t/2}v^{-(s+t)/2}\mu(s,t) \mm(s,t)\,,
\ee
where $\mm(s,t)$ is the Mellin amplitude and
\begin{eqnarray}
\mu(s,t)&=&\G\bigg(\frac{\D_{34}-s}{2}\bigg)\G\bigg(-\frac{\D_{12}+s}{2}\bigg)\G\bigg(\frac{s+t}{2}\bigg)\G\bigg(\frac{s+t+\D_{12}-\D_{34}}{2}\bigg)\nonumber\\
&&\G\left(\frac{\D_1+\D_2-t}{2}\right)\G\left(\frac{\D_3+\D_4-t}{2}\right),
\end{eqnarray} 
is the measure with $\D_{ij}=\D_i-\D_j$. The Mellin amplitude admits a partial wave decomposition \cite{Mack:2009mi},
\be\label{mell}
\mm(s,t)=\sum_{J=0}^\infty \int_{-\infty}^{\infty} d\n b_J(\n^2)\g(\n,t)\g(-\n,t)\zeta({\D_i,t})\mathcal{P}_{\n,J}(s,t,\{\D_i\})\,,
\ee
where $\mathcal{P}_{\n,J}(s,t)$ is the Mack polynomial;
\be
\g(\n)=\frac{\G(\frac{\D_1+\D_2+J+i\n-h}{2})\G(\frac{\D_3+\D_4+J+i\n-h}{2})\G\left(\frac{h+i\n-J-t}{2}\right)}{\sqrt{8\pi}\G(i\n)}\,,\ee
and
\be\label{tfactor}
\zeta({\D_i,t})=\frac{1}{\G(\frac{\D_1+\D_2-t}{2})\G(\frac{\D_3+\D_4-t}{2})}.\ee
This will be the focal point of our analysis.
We consider the $t$-channel decomposition with $\Delta_1=\Delta_4$ and $\Delta_2=\Delta_3$. In Appendix C of \cite{Costa2012a}, it was shown that the Regge limit of Mellin amplitude matches with the usual momentum space Regge limit\footnote{In the position space, the Regge limits correspond to a specific kinematic configuration of the four operators in the Lorentzian signature \cite{Cornalba:2007fs, Cornalba:2008qf}.}. In this work, we are however interested in the conformal Regge limit of the Mellin amplitude, irrespective of the physical implications in the momentum space. For large $s$ and fixed $t$, the Mack polynomial takes the form \cite{Costa2012a},
\be
\lim_{s\rightarrow\infty}\mathcal{P}_{\n,J}(s,t)=s^J a_J\,, \ \text{where}\ a_J=\frac{ (2-h-i\n+J)_J(2-h+i\n+J)_J}{(h+i\n-1)_J(h-i\n-1)_J}\,. \label{aJ}
\ee
The factor $a_J$ becomes $1$ for general $\n$ and integer $J$. \eqref{mell} can be separated in terms of even and odd spins, 
\be
\mm(s,t)=\mm_{+}(s,t)+\mm_{-}(s,t)\,,
\ee
\begin{figure}[!htb]
  \centering
   \begin{tikzpicture}[scale=1]
   \draw[-{stealth},
   line width=1mm,
   color=black] (-2,0)--(7,0);
   \draw[-{stealth},
   line width=1mm,
   color=black] (0,-2)--(0,6);
   \filldraw  (0,0)circle[radius=4pt]
   (1.5,0)circle[radius=4pt]
   (3,0)circle[radius=4pt]
   (4.5,0)circle[radius=4pt]
   (6,0)circle[radius=4pt];
   \draw
   [
   line width=1mm,
   color=magenta,
   decoration={markings, mark=at position 0.50 with {\arrow[line width=2mm]{stealth}}},
   postaction={decorate}
   ]
   (6.5,0.4)--(0,0.4); 
   \draw
   [
   line width=1mm,
   color=magenta,
   ] 
   (0,0.4) arc[start angle=90, end angle=270, radius=0.4cm];   
   \draw
   [
   line width=1mm,
   color=magenta,
   decoration={markings, mark=at position 0.60 with {\arrow[line width=2mm]{stealth}}},
   postaction={decorate}
   ]
   (0,-0.4)--(6.5,-0.4); 
   \draw
   [
   line width=1mm,
   color=ForestGreen,
   decoration={markings, mark=at position 0.60 with {\arrow[line width=2mm]{stealth}}},
   postaction={decorate}
   ]
   (-0.8,5.7)--(-0.8,-2);    
   \filldraw  (2,3)circle[radius=4pt];
   \draw[
   color=ForestGreen,
   line width=1mm,
   decoration={markings, mark=at position -0.3  with {\arrowreversed[scale=1,ForestGreen]{>}}},
   postaction={decorate}
   ] (2,3)circle[radius=10pt];
   \draw (-0.2,0.7) node {$0$};
   \draw (1.5,0.7) node {$2$};
   \draw (3,0.7) node {$4$};
   \draw (4.5,0.7) node {$6$};
   \draw (6,0.7) node {$8$};
   \draw
   [
   ultra thick,
   color=magenta
   ] (6,-0.9) node {\huge$C$};
   \draw
   [
   ultra thick,
   color=ForestGreen
   ] (-1.5,5) node {\huge$C'$};
   \draw (6,5.1) node {\huge$J$};
   \draw[color=MediumOrchid] (3.2,3.5) node {\huge$j(\nu)$};
   \draw[very thick] (5.7,4.7)--(6.3,4.7);
   \draw[very thick] (5.7,4.7)--(5.7,5.45);
   \end{tikzpicture} 
   \caption{Contour for SW transform}\label{SWcontour}
\end{figure}
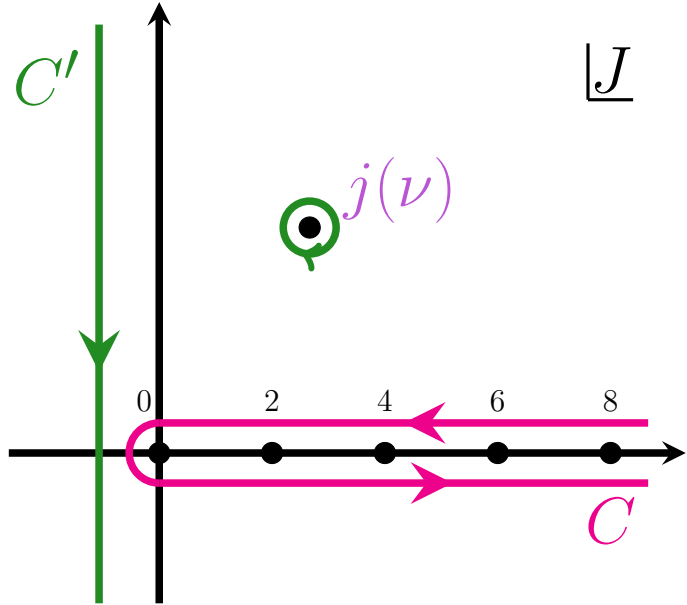 
where $\pm$ respectively stands for even and odd spins and,
\be
\mm_{\pm}(s,t)=\frac{1}{2}\sum_{J=0}^\infty \int d\n b^\pm_J(\n^2) \g(\n,t)\g(-\n,t)\zeta({\D_i,t}) s^J[1\pm(-1)^J]\,.
\ee
Next,  using the Sommerfeld-Watson (SW) transform, we replace $\sum_J$ in terms of a complex integral along the contour$-C$ (in fig.\eqref{SWcontour}),
\be
\sum_{J}\equiv \frac{1}{2\pi i}\oint_C dJ\frac{\pi e^{i\pi J}}{\sin\pi J}\,.
\ee
which picks up only integer poles in $J$. Recalling that $+$ and $-$ signs stand for contributions from even and odd spin respectively we have the following expressions,
\be\label{mel}
\mm_\pm(s,t)=\frac{1}{2\pi i} \oint dJ \frac{\pi}{\sin\pi J} \int d\n \zeta({\D_i,t})\g(\n,t)\g(-\n,t) s^J e^{i\pi J/2} \begin{cases} b^+_J(\n^2)\cos \pi J/2\,, +\\ - ib^-_J(\n^2)\sin \pi J/2\,, -\end{cases}\,.
\ee 
Following \cite{Costa2012a}, we analytically continue $J$ from integer to complex values, {\it i.e.} deform the contour $C\rightarrow C'$ (see figure \eqref{SWcontour}), to pick up the poles $J=J(\n)$ in the complex plane. The poles of $J=J(\nu)$ are determined from the spectral function $b_J(\nu)$\footnote{The leading Regge trajectory is determined by the largest exponent of $s^{J(\n)}$.}. From \cite{Costa2012a}, we make the correct identification of the spectral function for the fishnet CFT. 
\be
b_J(\nu^2)\frac{i\nu}{2\pi K_{2+i\nu,J}} = \frac{1}{c_2(\nu, J)}\frac{\left(E_{\Delta,J}^{(n)}\right)^p}{1-\chi E^{(n)}_{\Delta,J}}\,,
\ee
where $p=1,2,1$ for respectively $n=0,1,2$-magnon graphs. Putting in the normalizations,
\begin{align}
\begin{split}
K_{\D,J}=&
\frac{
	4^{1-J} 
	\Gamma (-h+\Delta +1) 
	\Gamma (J+\Delta )
	(\Delta -1)_J
}
{
	\Gamma \left(\frac{J+\Delta -\Delta _{12}}{2}\right) 
	\Gamma \left(\frac{J+\Delta +\Delta _{12}}{2}\right)
	\Gamma \left(\frac{J+\Delta -\Delta _{34}}{2}\right)
	\Gamma \left(\frac{J+\Delta +\Delta _{34}}{2}\right)	  
}\\
&\times \frac{1}
{
	\Gamma \left(\frac{J-\Delta +\Delta _1+\Delta _2}{2}\right) 
	\Gamma \left(\frac{J-\Delta +\Delta _3+\Delta _4}{2}\right)
	\Gamma \left(\frac{\Delta +\Delta _1+\Delta _2+J-2h}{2}\right) 
	\Gamma \left(\frac{\Delta +\Delta _3+\Delta _4+J-2h}{2}\right)},
\end{split}
\end{align}
and,
\be\label{3.12}
\begin{split}
	c_2(\n,J)=\frac{
		2 \pi ^{2 h+1} (-1)^J
		\Gamma (J+1)
		\Gamma (\Delta -h)
		\Gamma (J+\Delta -1)
		\Gamma \left(\frac{2 h+J-\Delta -\Delta _{12}}{2}\right)
		\Gamma \left(\frac{2 h+J-\Delta +\Delta _{12}}{2}\right)
	}
	{
		\Gamma (\Delta -1)
		\Gamma (h+J) 
		\Gamma \left(\frac{J+\Delta -\Delta _{12}}{2}\right)
		\Gamma \left(\frac{J+\Delta +\Delta _{12}}{2}\right)
		\Gamma (2 h+J-\Delta )
	},
\end{split}
\ee 
we get,
\begin{shaded}
	\begin{align}\label{melfn}
	\begin{split}
	\mm^\pm(s,t)&=\frac{1}{2\pi i} \oint dJ \frac{\pi}{\sin\pi J} \int_{-\infty}^\infty d\n \left(\frac{s}{4}\right)^J e^{i\pi J/2}\nu \sinh \pi\nu \,\zeta({\D_i,t})\\
	&\times\frac{
		(-1)^{-J}  (J+1)    
		\Gamma (J-i \nu +2)
		\Gamma (J+i \nu +2) 
		\Gamma \left(\frac{-J-t-i \nu +2}{2}\right) \Gamma \left(\frac{-J-t+i \nu +2}{2}\right)
					}
	{
		2\pi ^6 
		\Gamma \left(\frac{J-\Delta_{12}-i \nu +2}{2}\right) 
		\Gamma \left(\frac{J+\D_{12}-i \nu +2}{2}\right) 
		\Gamma \left(\frac{J-\Delta_{34}+i \nu +2}{2}\right) 
		\Gamma \left(\frac{J+\D_{34}+i \nu +2}{2}\right)
			}\, \frac{\left(E_{\Delta,J}^{(n)}\right)^p}{1-\chi_n E^{(n)}_{\Delta,J}} P_J^{\pm}, 
	\end{split}
	\end{align} 
\end{shaded}	
where,
\be
P_J^{\pm}=\begin{cases} \cos \pi J/2\,, +(\text{even spin})\\ - i\sin \pi J/2\,, -(\text{odd spin})\end{cases}\,.
\ee
is the phase factor associated with the even and odd parts. We note that for the zero-magnon case, with $\Delta_i=1$, the t-independent part of the amplitude in \eqref{melfn} exactly matches with the momentum space amplitude for $0-$magnon correlator in \cite{Korchemsky2018}.\ (We have put $\D=2+i \nu$ while in \cite{Korchemsky2018}, the author uses $\D=2+2i\nu$. The agreement of the two expressions assumes that this issue has been taken care of) . \footnote{One has to take the $z\rightarrow \infty$ of eqn 4.21 in \cite{Korchemsky2018}.}. 

Note that the term $e^{J\p/2}P_{J}^{\pm}$ takes care of $(s\rightarrow -s)$ in the SW transform and from now on we will dispense with this term by writing out the $(s\rightarrow -s)$ term separately. In the following sections we compute the Regge limit of Mellin amplitudes for 0,1 and 2-Magnon correlators.
\section{$0-$magnon correlator}\label{0mag}
In this section we will obtain the Regge limit of the 0-magnon correlator in Mellin space. The Regge limit of the scattering amplitude has already been analyzed in \cite{Korchemsky2018}. We perform similar analysis in Mellin space as a warm up for the other magnon graphs. Upto some $t$-dependent factors, we obtain a match with the Regge amplitude computed in \cite{Korchemsky2018}. For the $0-$magnon correlator, the external operator dimensions are $\D_1=\D_2=\D_3=\D_4=1$. The Mellin amplitude in the Regge limit is given by \eqref{melfn},
	\begin{align}\label{mel0}
	\begin{split}
	\mm_{(0)}^\pm(s,t)=&\Bigg[\frac{\pm 1}{2\pi i} \oint dJ \frac{\pi}{\sin\pi J} \int_{-\infty}^\infty d\n (s/4)^J \nu \sinh \pi\nu~ \zeta_0({\D_i,t})\\
	&\times \frac{ (J+1)  \Gamma (J-i \nu +2) \Gamma (J+i \nu +2) \Gamma \left(\frac{-J-t-i \nu +2}{2}\right) \Gamma \left(\frac{-J-t+i \nu +2}{2}\right)}{2\pi ^6 \Gamma \left(\frac{J-i \nu +2}{2}\right)^2 \Gamma \left(\frac{J+i \nu +2}{2}\right)^2}\frac{E_{2+i\n,J}^{(0)}}{1-\c_0 E_{2+i\n,J}^{(0)} }\Bigg] \\
& \pm(s\rightarrow -s),
	\end{split}
	\end{align} 
\noindent
where, from \eqref{eigs} we have the for the $0-$Magnon amplitude,
\be 
E_{\D,J}^{(0)}=\frac{16\pi^4 c^4}{(J+2-\D)(J+4-\D)(J+\D)(J+\D-2)},\qquad \chi_0=(16\pi^2\xi^2)^2,
\ee 
and $\zeta_0({\D_i,t})=\G(1-\frac{t}{2})^{-2}$. Also note that the extra sign in front of the Mellin amplitude stems as discussed following  \eqref{melfn}. 
Putting $\D=2+i\n$ above we obtain the following expression\footnote{Note that the authors of \cite{Korchemsky2018, Kazakov:2018hrh} use $\Delta=2+2 i \n$.},
\begin{align} \label{4.3}
\begin{split}
\frac{E_{2+i\n,J}^{(0)}}{1-\c_0 E_{2+i\n,J}^{(0)} }&=\frac{16\pi^4 c^4}{\left(J^2+\n^2\right)\left((J+2)^2+\n^2\right)-4f^4}
\end{split}
\end{align} 
with $f=4\sqrt{2}c\p^2\x$. 
 \subsection{Regge limit}
Solving for the poles of \eqref{4.3}, we obtain the Regge trajectories,  
\begin{align}\label{Regge0}
\begin{split}
J^\pm_2=-1+\sqrt{1-\n^2\pm 2\sqrt{f^4-\n^2}}\,,\ J^\pm_4=-1-\sqrt{1-\n^2\pm 2\sqrt{f^4-\n^2}}\,.
\end{split}
\end{align}
The leading Regge trajectories come from the pole(s) having the most positive real part (in the limit $s\rightarrow \infty$). Thus the leading Regge trajectory is obtained from $J_2^+$ \cite{Korchemsky2018}. The integral over $\nu$ in \eqref{mel0} is performed as follows. We first compute residue of the spectral function due to the Regge poles. Schematically this is given by, 
\be
\text{Res.}\bigg[\frac{E_{\Delta,J}}{1-\chi_0 E_{\Delta,J}}\bigg]_{J=J_i}=\frac{4c^4\p^4}{\left(J_i+1\right)\left(J_i\left(J_i+2\right)+\n^2\right)}\,,
\ee
where the residue is evaluated at the Regge poles $J_i=J_{2}^{\pm}$. Evaluating the residue around the Regge poles (for leading Regge trajectories), the Mellin amplitude is given by\footnote{We are just looking at the even spin hence considering $\mm_{(0)}^+$. The odd spin case i.e, $\mm_{(0)}^-$ can be tackled in a same fashion by putting proper signs as delineated in the discussion following \eqref{melfn}.}, 
\be \label{mel0Regge1}
\mm_{(0)}^{+}=\frac{2c^4}{\pi}\,\zeta_0({\D_i,t})\sum_{J_2^+,J_2^-}\int_{-\infty}^\infty d\n \left(\frac{s}{4}\right)^J\, F(\n,J)+\left(s \rightarrow -s\right),
 \ee
where, 
\be
F(\n,J)=\frac{ \nu   \sinh (\pi  \nu ) \Gamma (J-i \nu +2) \Gamma (J+i \nu +2) \Gamma \left(\frac{-J-t-i \nu +2}{2}\right) \Gamma \left(\frac{-J-t+i \nu +2}{2}\right)}{\sin (\pi  J) \left(J (J+2)+\nu ^2\right) \Gamma^2 \left(\frac{J-i \nu +2}{2}\right) \Gamma^2 \left(\frac{J+i \nu +2}{2}\right)}
. \ee
We will now evaluate this integral in weak coupling limit, $f\rightarrow 0$ and strong coupling limit, $f\rightarrow \infty$.
\subsection{Weak Coupling: $f\rightarrow 0$}
\paragraph{} Following \cite{Korchemsky2018}, we manipulate the integral in \eqref{mel0Regge1} into a form that is valid for primarily weak coupling and then we evaluate the integral in the weak coupling limit.
This integral can be effectively reduced to an integral over the interval $-f^2\leq\n\leq f^2$ so that,  
\be\label{zero1}
\mm_{(0)}^+(s,t)\approx\frac{2c^4}{\pi}\,\zeta_0({\D_i,t})\int_{-f^2}^{f^2} d\n [(s/4)^{J_2^+}F^+(\n)-(s/4)^{J_2^-}F^-(\n)]\, +\left(s \rightarrow -s\right), 
\ee
with,
\be 
F^{\pm}(\n)=F(\n,J_{2}^{\pm}(\n))
,\ee 
where the approximate sign denotes  that this equality is valid modulo terms of order $O(s^{-1})$ which vanish in the limit $s\rightarrow\infty$ \footnote{We thank Gregory Korchemsky for sharing his notes on this manipulation with us. Interested readers will find the details of this manipulation in Appendix \ref{poleanalysis0mag}}. It is convenient to perform a change of variables $\n=f^2\sqrt{1-x^2}$ and  define, 
\begin{align}
\begin{split}
j(x)&=J_2^+/f^2=\left(-1+\sqrt{1+2f^2 x+f^4(x^2-1)}\right)/f^2, \\
\f(x)&= F(f^2\sqrt{1-x^2},f^2j(x)),
\end{split}
\end{align}
and in this notation, $J_2^-=f^2 j(-x)$ and $F^{\pm}(\n)=\f(\pm x)$. Introducing, $q=f^2\log (s/4)$, we can finally write,
\begin{align}\label{mel0int}
\begin{split}
\mm_{(0)}^+(s,t)&=\frac{2c^4f^2}{\pi}\,\zeta_0({\D_i,t})\int_0^1 \frac{x dx}{\sqrt{1-x^2}} \left\lbrace \f(x)e^{qj(x)}-\f(-x)e^{qj(-x)}\right\rbrace+(s\rightarrow -s),\\
&=\frac{2c^4f^2}{\pi}\,\zeta_0({\D_i,t})\int_{-1}^1 \frac{x dx}{\sqrt{1-x^2}} \f(x)e^{qj(x)}+(s\rightarrow -s)\,,
\end{split}
\end{align}
where in the last line, we have performed a change of variables $x \rightarrow -x$ to combine the two regions of integration. Now we will analyze the Regge amplitude in the weak coupling limit. More precisely, we take the following set of limits.
$$f\rightarrow 0,\qquad s\rightarrow \infty, \qquad q=f^2 \log \left(\frac{s}{4}\right) \rightarrow \text{fixed}.$$ Expanding the integrand in the weak coupling limit, we write first few terms,
\begin{align} \label{0mfr1}
\begin{split}
&\frac{x }{\sqrt{1-x^2}} \f(x)e^{qj(x)}\\
=&\sqrt{1-x^2} e^{q x} \Gamma \left(1-\frac{t}{2}\right)^2\Bigg[\frac{1}{2 x}
-\frac{f^2 }{4 x^2} \left(x (q-4 x)+2 x^2 \psi ^{(0)}\left(1-\frac{t}{2}\right)-1\right)\\
&\,+\frac{f^4}{48 x^3}
\bigg\lbrace 3 q^2 x^2+6 x^2 \bigg(2 x\, \psi ^{(0)}\left(1-\frac{t}{2}\right) \left(q+x \psi ^{(0)}\left(1-\frac{t}{2}\right)-4 x\right)\\
&+\left(2 x^2-1\right) \psi ^{(1)}\left(1-\frac{t}{2}\right)\bigg)-6 q \left(2 x^3+x\right)+2 \left(2 x^2+1\right) \left(\pi ^2 x^2+3\right)\bigg\rbrace+O(f^5).
\end{split}
\end{align}
This integral can be done with the help of integrals described in  Appendix \ref{intweak} and specifically, the integrals that go into the final evaluation are those in \eqref{nneg}. Upto a few orders of expansion in $f$ we have the following result,
\begin{shaded}
	\begin{align}\label{0mfr}
	\begin{split}
	\mm_{(0)}^+(s,t)&=2c^4f^2 \bigg[
	\frac{1}{4} (q (\pi  \pmb{L}_1(q)+2) I_0(q)-(\pi  q \pmb{L}_0(q)+2) I_1(q))-f^2 \bigg\lbrace\frac{  I_1(q)}{2q}\left[\psi^{(0)}\left(1-\frac{t}{2}\right)-2\right]\\
	&\hspace{1.5 cm}+\frac{q}{8}  (q (\pi  \pmb{L}_1(q)+2) I_0(q)-(\pi  q \pmb{L}_0(q)+2) I_1(q))-\frac{1}{8} \left(\left(\pi  q^2 \pmb{L}_1(q)+2 q^2-2\right) I_0(q)\right.\\
&\left.\hspace{4 cm}-L (\pi  q \pmb{L}_0(q)+2) I_1(q)+2\right)\bigg\rbrace+O(f^4)\bigg]+(s\rightarrow -s).
	\end{split}
	\end{align}
\end{shaded}
This is the main result in the weak coupling limit of $0-$magnon correlator. Apart from the $t-$dependent factors, the integrand arranges itself into the same structure as that of \cite{Korchemsky2018}. We are computing Regge amplitudes from the {\it Conformal Regge theory} (CRT) point of view, independent of the LSZ approach in \cite{Korchemsky2018}. The CRT also aids to compute the Regge limit of the 1-magnon correlators with off-shell states.
\subsubsection{Comparison with existing results for $0$-magnon}\label{cwer0m}
We now discuss the differences between the Mellin amplitude in \eqref{0mfr} and the final results of the momentum space computations in \cite{Korchemsky2018}. There are two basic points of difference with regard to $t$-dependent and independent factors. 
	  \begin{itemize}
	\item {\it $t$-dependent factors}: Firstly, comparing \eqref{mel0Regge1} and the equivalent eqpression from \cite{Korchemsky2018} (see eq (5.5) of \cite{Korchemsky2018}), we observe that upto overall factors, the term in \eqref{mel0Regge1} has an extra $t$-dependent factor $$\Gamma \left(\frac{-J-t-i \nu +2}{2}\right) \Gamma \left(\frac{-J-t+i \nu +2}{2}\right)$$
	It is this factor which in the weak coupling limit gives rise to the additive terms $\propto (\psi^{(n)}\left(1-\frac{t}{2}\right))^m$ in \eqref{0mfr1} ( and hence \eqref{0mfr}). These factors were not present in the final result of \cite{Korchemsky2018}. We can also shed some light on the  conceptual origin of this discrepancy. Note that while the author of \cite{Korchemsky2018} had worked in momentum space, we are working in Mellin space, which is auxiliary to usual momentum space and these terms occur naturally in this formalism. We don't have a deeper understanding of this issue and leave this as a conjecture that in order to get the momentum space Regge amplitude from the {\it Conformal Regge theory} (CRT) techniques, we have to throw away the terms $\propto (\psi^{(n)}\left(1-\frac{t}{2}\right))^m$ in the final Mellin amplitude.
	\item {\it $t$-independent factors}: Secondly, in general we have used $\Delta= 2+ i \nu$ while authors of \cite{Korchemsky2018} ( as well as \cite{Gromov2018}) have consistently used $\Delta= 2+ 2i\nu$ for explicit computations. This leads to a difference between the $t$-independent part of \eqref{mel0} and it's equivalent, eq (5.5) of \cite{Korchemsky2018}. Note that however, as a function of $\Delta$, the $t$-independent parts are equal. Therefore, although the $t$- independent factors of the final result  \eqref{0mfr} differs from that of \cite{Korchemsky2018} by some numerical factors (both overall and relative), the $f$ and $\log L$ dependence of the Mellin amplitude remain unaffected. This is simply a choice of convention and once we take this into account, we reproduce the exact Regge amplitude ( i.e exact w.r.t relative and overall numerical coefficients) reported in \cite{Korchemsky2018}.        
\end{itemize}
Once we take care of these issues (i.e, put $\D=2 + 2i \nu$ from the beginning and ignore terms $\propto (\psi^{(n)}\left(1-\frac{t}{2}\right))^m$ in the final Mellin amplitude), \eqref{0mfr} matches with results of \cite{Korchemsky2018}.

\subsection{Strong Coupling: $f\rightarrow\infty$} 
\paragraph{} For strong coupling, we perform an order of magnitude analysis \cite{Korchemsky2018}. The procedure is as follows, we first look at the behavior of the Regge poles $J_2^{\pm}$ and $J_4^{\pm}$
as a function of $\nu$. We observe that the dominant contribution comes from $J_2^+$ near $\n=0$\footnote{All other poles are subleading near $\n=0$ and all the poles including $J_2^+$ are subleading in the limit $\nu \rightarrow \infty$},
\be 
J_{2}^{\pm}=-1\pm \sqrt{2 f^2+1}\mp \frac{\left(f^2+1\right) \nu ^2}{2 f^2 \sqrt{2 f^2+1}}+O(\n^4).
\ee 
We clearly see that in the Regge limit, $J_{2}^+$ dominates over $J_{2}^-$ which is exponentially suppressed. Also note that this is true for any coupling. Let us define,
\be 
J_2^+=J_R^+ -\d \nu^2 + O(\n^4), \qquad J_R^+=-1+ \sqrt{2 f^2+1},\qquad \d=\frac{\left(f^2+1\right) }{2 f^2 \sqrt{2 f^2+1}}
.\ee
Hence around  $\n=0$ from \eqref{mel0Regge1} we have for $\mm^+_{(0)}$,
\begin{align}
\begin{split}
\mm^+_{(0)} &\approx 2c^4 \,\zeta_0({\D_i,t}) (s/4)^{J_R^+-\d\n^2}\n^2\frac{\G(J_R^++2)^2\G\left(\frac{-J_R^+-t+2}{2}\right)^2}{\sin(\p J_R^+)(J_R^+(2+J_R^+))\G\left(\frac{J_R^++2}{2}\right)^4} \,.
\end{split}
\end{align} 

Since the dominant contribution to the Regge amplitude ( i.e in the limit $s \rightarrow \infty$) comes from the region $\nu\sim 0$, the $\nu$ integral effectively reduces to,
\be 
\int d\n \n^2 (s/4)^{-\d\n^2}\sim \frac{\sqrt{\pi } }{4 \delta ^{3/2} \log ^{\frac{3}{2}}(s)}.
\ee 
Further around $f\rightarrow\infty$,
\begin{align} 
\begin{split}
J_R^+=-1+\sqrt{2}f+O\left(\frac{1}{f}\right)\,, \ \d= \frac{1}{2\sqrt{2}f}+O\left(\frac{1}{f^2}\right)\,.
\end{split}
\end{align} 
Collecting everything, we obtain the Regge amplitude in the strong coupling to be,
\begin{shaded}
\be 
\mm_{(0)}^{+}(s,t)\sim \left[- 2\sqrt[4]{8}c^4\sqrt{\frac{f}{\p}}\csc(\sqrt{2}\p f) \,\frac{s^{\sqrt{2}f}}{s\log ^{\frac{3}{2}}(s)}\frac{\G\left(\frac{3-t-\sqrt{2}f}{2}\right)^2}{\G\left(1-\frac{t}{2}\right)^2}\right]+\,(s\rightarrow -s).
\ee 
\end{shaded}
A similar order of magnitude analysis can be done for the weak coupling also. We find that the leading behavior matches one obtained from \eqref{mel0int}. 

\section{$1-$magnon correlator}\label{1mag}
For the $1$-magnon correlator, we put $\Delta_1=\Delta_4=2, \Delta_2=\Delta_3=1$ in \eqref{melfn} so that, 
	\begin{align}\label{mel1}
	\begin{split}
	\mm^{\pm}_{(1)}(s,t)=\Bigg[& \frac{\pm1}{2 \pi i}\oint dJ \frac{\pi}{\sin\pi J} \int_{-\infty}^\infty d\n\, s^J e^{i\pi J/2}\nu \sinh \pi\nu ~\zeta_1({\D_i,t})\\
	&\times\frac{
		2 (J+1)  \Gamma \left(\frac{J-i \nu +2}{2}\right)
		\Gamma\left(\frac{J+i \nu +2}{2}\right)
		\Gamma \left(\frac{2-J-t-i \nu }{2}\right)
		\Gamma \left(\frac{2-J-t+i \nu }{2}\right)
	}
	{
		\pi ^7 
		\Gamma \left(\frac{J-i \nu +1}{2}\right)
		\Gamma \left(\frac{J1i \nu +1}{2}\right)
	} \frac{\left(E_{2+i\n,J}^{(1)}\right)^2}{1-\chi_1 E_{2+i\n,J}^{(1)}}\Bigg]\pm(s\rightarrow -s), 
	\end{split}
	\end{align}
\noindent
where, $\zeta_1({\D_i,t})=\G(\frac{3-t}{2})^{-2}$. The spectral function for the $1-$magnon case is given by \eqref{eigs}, 
\be 
E^{(1)}_{\D,J}=(-1)^J\frac{4\pi^2 c^2}{(J+\D-1)(J-\D+3)}, \qquad \chi_1=16\pi^2\xi^2,
\ee 
and thereby,
\begin{align} \label{Regge1}
\begin{split}
\frac{(E^{(1)}_{\D,J})^2}{1-\chi E^{(1)}_{\D,J}}=\frac{(4\pi^2c^2)^2}{(J+\D-1)(J-\D+3)((J+\D-1)(J-\D+3)-(-1)^J g^2)}\,.
\end{split}
\end{align}
where $g=2\pi c\sqrt{\chi_1}=8 \pi ^2 c \xi$. We now determine the Regge poles for this spectral function. Replacing $\D=2+i\n$ in \eqref{Regge1}, we can see that, the above has four sets of poles at,
\be\label{Reggepol1mag}
J=\begin{cases}-1\pm i\n\\
-1\pm i\n\end{cases}\,, \ \ J=\begin{cases} -1\pm \sqrt{g^2-\n^2}\,, \ J={\rm even}\\ -1\pm i\sqrt{g^2+\n^2}\,, J={\rm odd}\end{cases}\,.
\ee
There are a few observations in order. The leading trajectory clearly comes from $J=-1 + \sqrt{g^2-\nu^2}$. For $g=0$, the first and second set above collide to give double poles.
\subsection{Regge limit: even spin}
\paragraph{} In this section we compute the Regge amplitude for the even spin. For even spin we need to consider $\mm^{+}_{(1)}$. Regge poles are at  $$J^{\pm}_{e}=-1\pm\sqrt{g^2-\n^2}$$ Also we evaluate
\be 
\text{Res.}\left[\frac{\left(E_{2+i\n,J}^{(1)}\right)^2}{1-\chi_1 E_{2+i\n,J}^{(1)}}\right]_{J=J_e^{\pm}}=\frac{8c^4\pi^4}{g^2(J_e^{\pm}+1)}.
\ee 
So that, after the $J-$integral the Mellin amplitude can be written as (where we have dispensed with the factor $(-1)^J\,P_J^{\pm}$ as in the $0$-Magnon analysis ),
\be\label{integral1mag}
\mm^{+}_{(1)}(s,t)=\zeta_1({\D_i,t})\int_{-\infty}^\infty d\n\left[s^{J_e^+}F(J_e^+)+s^{J_e^-}F(J_e^-)\right]+(s\rightarrow -s),
\ee
with, 
\be
F(J_e^{\pm})=\,\frac{	16 c^4 \nu \sinh (\pi\nu )}{\pi^2g^2\sin (\pi  J_e^{\pm})} \,
\frac{
	\Gamma \left(\frac{J_e^{\pm}-i \nu +2}{2}\right) 
	\Gamma \left(\frac{J_e^{\pm}+i \nu +2}{2}\right)
	\Gamma \left(\frac{2-J_e^{\pm}-t-i \nu }{2}\right)
	\Gamma \left(\frac{2-J_e^{\pm}-t+i \nu }{2}\right)
}
{	
	\Gamma \left(\frac{J_e^{\pm}-i \nu +1}{2}\right)
	\Gamma \left(\frac{J_e^{\pm}+i \nu +1}{2}\right)
}.
\ee
Again it is shown in the Appendix \ref{poleanalysis1mag}, that $\n$ integral in \eqref{integral1mag} reduces effectively to an integral over the range $[-g,g]$ as in the $0$-magnon case,\\
\be\label{1even}
{\cal{M}}_{(1)}^{+}(s,t) =\zeta_1({\D_i,t})\int_{-g}^{g} d\n \left( F(J_e^+)s^{J_e^+}- F(J_e^-)s^{J_e^-} \right) \,.
\ee
We will use this expression to investigate the weak coupling $g\rightarrow 0$ limit.

\subsubsection{Weak Coupling: $g\rightarrow 0$}
\paragraph{ } In order to evaluate \eqref{1even}, we use the following transformation of variables, 
\begin{eqnarray}
\n=g\sqrt{1-x^2}\,,\ j(\pm x)=J^{\pm}_{e}/f=(-1\pm gx)/g\,,\ F(\pm x)=F(-1\pm g x, g\sqrt{1-x^2})\,,
\end{eqnarray}
and rewrite the Mellin amplitude as, 
\begin{align}\label{int1e}
\begin{split}
\mm^{+}_{(1)}(s,t)&=2g\zeta_1({\D_i,t})\int_0^1 \frac{x dx}{\sqrt{1-x^2}}\bigg[F(x)e^{qj(x)}-F(-x)(s/4)^{qj(-x)}\bigg]+\,(s\rightarrow-s),\\
&=2g\zeta_1({\D_i,t})\int_{-1}^1 \frac{x dx}{\sqrt{1-x^2}}F(x)e^{qj(x)}+\,(s\rightarrow-s),\\
\end{split}
\end{align}
with $q=g\log s$ and
\be 
F(x)=-\frac{16c^4}{g\pi^2}\,\sqrt{1-x^2}\sinh\left(g\p \sqrt{1-x^2}\right)\csc(g\p x)\,\t\left(\frac{g}{2}(x;i\sqrt{1-x^2})\left|\frac{3-t}{2}\right.\right),\ee
with
\begin{align}\label{theta}
\begin{split}
\t(\a(a;b)|c)=\frac
{
	\Gamma \left(\a(a+b)+\frac{1}{2}\right)
	\Gamma \left(\a(a-b)+\frac{1}{2}\right)
}
{
	\Gamma \left(\a(a+b)\right) 
	\Gamma \left(\a(a-b)\right)
}\, \G(c-\a(a+b))\G(c-\a(a-b)).
\end{split}
\end{align} 
Analogous to the $0$-magnon case, we evaluate the integral \eqref{int1e} in the limit $$s \rightarrow \infty, \qquad g\rightarrow 0,\qquad q=g\log s \rightarrow \text{constant}.$$
The integrand now takes the form, 
\begin{align}
\begin{split} 
&2g\frac{x }{\sqrt{1-x^2}}F(x)e^{qj(x)}\\
&=-\frac{8c^4g^2}{s}\Gamma \left(\frac{3}{2}-\frac{t}{2}\right)^2e^{q x} \bigg[\frac{ \sqrt{1-x^2} }{\pi }
-\frac{ g  \sqrt{1-x^2} }{\p} \,x\left(\psi ^{(0)}\left(\frac{3}{2}-\frac{t}{2}\right)+\log (4)\right)\\
&\hspace{2.5 cm}+\frac{ g^2\sqrt{1-x^2} }{12\p}  \bigg(6 x^2 \psi ^{(0)}\left(\frac{3}{2}-\frac{t}{2}\right)^2+\left(6 x^2-3\right) \psi ^{(1)}\left(\frac{3}{2}-\frac{t}{2}\right)\\
&\hspace{2.5 cm}+12 x^2 \log (4) \psi ^{(0)}\left(\frac{3}{2}-\frac{t}{2}\right)+2 \pi ^2 x^2+6 x^2 \log ^2(4)+\pi ^2\bigg)\bigg]+O(g^5).\\
\end{split}
\end{align}
Again as in the case of zero magnon weak coupling, we can do the integration term by term. Effectively it gets reduced to evaluating the following integral,
\be 
\int_{-1}^{1}dx\, \sqrt{1-x^2} e^{q x}=\frac{\pi  I_1(q)}{q}.
\ee 
Further details of the integrals are explained in Appendix \ref{intweak}, especially the integrals that go into this evaluation are effectively those in \eqref{npos}. However here we write the explicit expression upto a few orders of expansion in $g$ taking into account $\zeta(\D_i,t)$,
\begin{shaded}\label{1meleve0}
	\begin{align}
	\begin{split}
	&\mm^{+}_{(1)}(s,t)= -\frac{8c^4g^2}{s} 
	\bigg[
	\frac{  I_1(q)}{q}
	-g\frac{  I_2(q)}{q}\left\lbrace\psi ^{(0)}\left(\frac{3}{2}-\frac{t}{2}\right)+\log (4)\right\rbrace+O(g^2)\bigg]+(s\rightarrow-s).
\end{split}
	\end{align}
\end{shaded}
This is the main result for weak coupling of the $1-$magnon correlator. 
 
\subsubsection{Strong Coupling: $g\rightarrow \infty$}
Similarly, for the strong coupling limit, the contribution to the integral occurs around $\nu \sim 0$ and is dominated by $J_e^+$,
\be 
J_e^+(\n)=-1+\sqrt{g^2-\n^2}=J_R-\frac{\n^2}{2g}-\frac{\n^4}{8g^3}+O(\n^6)\,, \ J_R=-1+g\,.
\ee 
Leading contribution to exponent of $s$ thus comes from the vicinity of $\n=0$, whereby the integrand of \eqref{integral1mag} becomes,
\be 
s^{J_e^+(\n)}F(J_e^+(\n))\approx\, s^{J_e^+(\n)} \frac{16c^4\n^2}{\p g^2\sin(\p J_R)}\frac{\G(1+J_R/2)^2\G(1-(J_R+t)/2)^2}{\G((J_R+1)/2)^2}
.\ee 
So the approximate integral is ,
\be 
\int d\n\, \n^2s^{J_e^+(\n)}\approx s^{J_R}\int d\n\, \n^2s^{-\frac{\n^2}{2g}} \sim g^{\frac{3}{2}}\sqrt{\frac{\p}{2}}\frac{s^{J_R} }{ \log ^{\frac{3}{2}}(s)} ,
\ee 
from which it follows that, 
\be 
\mm^+_{(1)}(s,t)\sim \frac{16c^4}{\p g^2\sin(\p J_R)}\frac{\G(1+J_R/2)^2\G(1-(J_R+t)/2)^2}{\G((J_R+1)/2)^2}\,g^{\frac{3}{2}}\sqrt{\frac{\p}{2}}\frac{s^{J_R} }{ \log ^{\frac{3}{2}}(s)} +\, (s\rightarrow-s).
\ee 
Further the strong coupling limit $g\rightarrow\infty$ we have the asymptotic relation $\frac{\Gamma \left(\frac{g+1}{2}\right)^2}{\Gamma \left(\frac{g}{2}\right)^2}\sim \frac{g}{2}$, and,
\begin{shaded}
\be 
\mm^+_{(1)}(s,t)\sim \left[-\frac{4\sqrt{2}c^4}{\sin(\p g)}\sqrt{\frac{g}{\pi }} \, \frac{s^{g} }{ s\log ^{\frac{3}{2}}(s)}\frac{\G\left(\frac{3-t-g}{2}\right)^2}{\G\left(\frac{3-t}{2}\right)^2}\right]+(s\rightarrow-s). 
\ee 
\end{shaded} 
The contribution from $J_{e}^-$ is exponentially suppressed compared to the above.
\subsection{Regge limit: odd spin}
\paragraph{} In this section we study the Regge amplitude corresponding to Regge poles associated with the SW transform for odd spins (see \eqref{Reggepol1mag}). Further we set $J^{\pm}_o=-1\pm i\sqrt{g^2+\n^2}$ and  we evaluate,
\be 
\text{Res.}\left[\frac{\left(E_{2+i\n,J}^{(1)}\right)^2}{1-\chi_1 E_{2+i\n,J}^{(1)}}\right]_{J=J^{\pm}_o}=-\frac{8c^4\pi^4}{g^2(J^{\pm}_o+1)}
.\ee
Therefore   after the $J$-integral, the Mellin amplitude is cast into the following form, 
\be \label{1odd}
\mm^{-}_{(1)}(s,t)=\zeta_1({\D_i,t})\int_{-\infty}^\infty d\n\left[s^{J^+_o}F(J^+_o)+s^{J^-_o}F(J^-_o)\right]-(s\rightarrow -s),
\ee 
with, 
\be
F(J^{\pm}_o)=\,\frac{	16 c^4 \nu \sinh (\pi\nu )}{\pi^2g^2\sin (\pi  J^{\pm}_o)} \,
\frac{
	\Gamma \left(\frac{J^{\pm}_o-i \nu +2}{2}\right) 
	\Gamma \left(\frac{J^{\pm}_o+i \nu +2}{2}\right)
	\Gamma \left(\frac{2-J^{\pm}_o-t-i \nu }{2}\right)
	\Gamma \left(\frac{2-J^{\pm}_o-t+i \nu }{2}\right)
}
{	
	\Gamma \left(\frac{J^{\pm}_o-i \nu +1}{2}\right)
	\Gamma \left(\frac{J^{\pm}_o+i \nu +1}{2}\right)
}.
\ee 
We evaluate this integral in both the strong coupling and weak coupling limit.
\subsubsection{Weak Coupling: $g\rightarrow 0$}
\paragraph{ } We study the weak coupling limit of \eqref{1odd}. 
To begin with, we define $\n^2+g^2=\hat{y}^2$ so that  we have  $J^{\pm}_o=-1\pm i\hat{y}$. With this change of variable \eqref{1odd}can be written as, 
\be 
\mathcal{M}_{(1)}^-=\zeta_1({\D_i,t})\int_{g}^\infty \frac{\hat{y}d\hat{y}}{\sqrt{\hat{y}^2-g^2}}\left[s^{J^+_o(\hat{y})}F(J^+_o(\hat{y}))+s^{J^-_o(\hat{y})}F(J^-_o(\hat{y}))\right]-(s\rightarrow -s),
\ee
with,
\begin{align}
\begin{split}
F(J^{\pm}_o(\hat{y}))=\frac
{16 c^4 \sqrt{\hat{y}^2-g^2}\, \sinh \left(\p\sqrt{\hat{y}^2-g^2} \right) \csc (\pm i\pi\hat{y})}{\pi^2  g^2}\,\t\left(\frac{1}{2}(\pm i\hat{y};i\sqrt{\hat{y}^2-g^2})\left|\frac{3-t}{2}\right.\right),
\end{split}
\end{align}\\
where we have used the definition in \eqref{theta} and,
\be 
\csc\left(\pi(-1\pm i x)\right)=- \csc(\pm i\pi x).
\ee 
Now we consider the transformation of the variable $\hat{y}=g/z$. We then have the integral as , 
\be 
\mathcal{M}_{(1)}^-=\,g~\zeta_1({\D_i,t})\int_{0}^1 \frac{dz}{z^2\sqrt{1-z^2}}\,\left[\left(\frac{s}{4}\right)^{J^+_o(z)}F(J^+_o(z))+\left(\frac{s}{4}\right)^{J^-_o(z)}F(J^-_o(z))\right] -(s\rightarrow -s),
\ee 
with, 
\begin{align}
\begin{split}
F(J^{\pm}_o(z))=\frac
{16 c^4 \sqrt{1-z^2}\, \sinh \left(\frac{\p g}{z}\sqrt{1-z^2} \right) \csc (\pm i\pi g/z)}{\pi^2 gz} \,\t\left(\frac{ig}{2z}(\pm 1;\sqrt{1-z^2})\left|\frac{3-t}{2}\right.\right).
\end{split}
\end{align}
Next we observe in above that,
\be 
F(J_{+}^o(-z))=F(J_{-}^o(z)).
\ee  
With this observation if we define,
\be 
j^o(\pm z)=-g^{-1}\pm iz^{-1} \qquad \mathcal{F}^o(\pm z)=F(J^{\pm}_o(z))\equiv F(gj^o(\pm z)),
\ee 
we can write (4.47) as 
\begin{align}
\mathcal{M}_{(1)}^-(s,t)&=\,g~\zeta_1({\D_i,t})\int_{0}^1 \frac{dz}{z^2\sqrt{1-z^2}}\,\left[e^{qj^o(z)}\mathcal{F}^o(z)+e^{qj^o(-z)}\mathcal{F}^o(-z)\right] -(s\rightarrow -s),\nonumber\\
&=\,g~\zeta_1({\D_i,t})\int_{-1}^{1}\frac{dz}{z^2\sqrt{1-z^2}}\,e^{qj^o(z)}\mathcal{F}^o(z) -(s\rightarrow -s),\nonumber\\
&=\,\frac{g}{s}~\zeta_1({\D_i,t})\int_{-1}^{1}\frac{dz}{z^2\sqrt{1-z^2}}\,e^{iq/z}\mathcal{F}^o(z) -(s\rightarrow -s).
\end{align}
with $ q=g\log s$. We evaluate this integral in the limit 
$$s \rightarrow \infty, \qquad g\rightarrow 0,\qquad q=g\log s \rightarrow \text{constant}.$$ 
As before $s\rightarrow \infty$ has already been taken into account in the Mellin amplitude. The weak coupling limit gives us the following expansion (upto order $O(g^3)$),
\begin{align}
\begin{split}
g\,\frac{e^{iq/z}\mathcal{F}^o(z)}{z^2\sqrt{1-z^2}}&=\frac{4c^4}{\pi}\Gamma \left(\frac{3}{2}-\frac{t}{2}\right)^2e^{iq/z}\sqrt{1-z^2}\left[ \frac{ i  g^2  }{ z^3}
+\frac{ g^3}{z^4}   \left\lbrace\psi ^{(0)}\left(\frac{3}{2}-\frac{t}{2}\right)+\log (4)\right\rbrace\right].
\end{split}
\end{align}  
Thus the problem is effectively reduced to the following integral with $n\in\mathbb{Z}^+$,
\be 
\int_{-1}^{1}dx\,e^{\frac{iq}{x}}\,\frac{\sqrt{1-x^2}}{x^n}=\frac{\pi ^2 e^{\frac{-i \pi  n}{2}} }{4} \left(\frac{2 \, _1\tilde{F}_2\left(\frac{n-2}{2};\frac{1}{2},\frac{n+1}{2};-\frac{q^2}{4}\right)}{\Gamma \left(2-\frac{n}{2}\right)}+\frac{ q\, _1\tilde{F}_2\left(\frac{n-1}{2};\frac{3}{2},\frac{n+2}{2};-\frac{q^2}{4}\right)}{\Gamma \left(\frac{3}{2}-\frac{n}{2}\right)}\right),
\ee
with,
\be 
_1\tilde{F}_2\left(a_1;b_1,b_2;z\right)=\frac{_1F_2\left(a_1;b_1,b_2;z\right)}{\G(b_1)\G(b_2)}.
\ee 
Thus we have the expression for Mellin amplitude,
\begin{shaded}\label{0magmelReggeweakfinal}
\be
\mm_{(1)}^-=-\frac{4c^4}{\pi s} \left[g^2\frac{ \pi  J_1(q)}{q}+ g^3   \left\lbrace\psi ^{(0)}\left(\frac{3}{2}-\frac{t}{2}\right)+\log (4)\right\rbrace\frac{\pi  J_2(q)}{q}\right]+O(g^4)-(s\rightarrow-s).
\ee
\end{shaded}	
We find that for large $q$
\be
\mathcal{M}_{(1)}^-(s,t) \sim \frac{\cos(\log s)}{s\log^{\frac{3}{2}} s}.
\ee
\subsubsection{Strong Coupling: $g\rightarrow \infty$} 
Analogous to the even spin case, maximal contribution again comes from $\n=0$\footnote{We note that the odd poles in \eqref{Reggepol1mag} contribute to exponential suppression via the term $\sin (\pi J_o^{\pm})$ for large $\nu$}. We write,
\be 
J_{\pm}^{o}=-1\pm ig\pm \frac{i\n^2}{2g}+O(\n^4)\,.
\ee 
Note that the coupling dependence gives a phase. And so we have to consider contribution from both $J_{+}^o$and $J_{-}^o$. For brevity, we set $J_{0\pm}=-1\pm ig$. Then,
\be
s^{J^\pm_o(\n)}F(J^\pm_o(\n))\approx \, s^{J^\pm_o(\n)} \frac{16c^4\n^2}{\p g^2\sin(\p J^\pm_o)}\frac{\G(1+J^\pm_o/2)^2\G(1-(J^\pm_o+t)/2)^2}{\G((J^\pm_o+1)/2)^2}\,.
\ee
And consequently,
\begin{align}
\begin{split} 
\int d\n\, \n^2s^{J^{+}_o(\n)}&\approx s^{J_{0+}}\int d\n\, \n^2s^{+\frac{i\n^2}{2g}}\sim + g^{\frac{3}{2}}\sqrt{\frac{\p}{4}}(1+i) \frac{is^{ig} }{s \log ^{\frac{3}{2}}(s)}\,,\\
\int d\n\, \n^2s^{J^{-}_o(\n)}&\approx s^{J_{0-}}\int d\n\, \n^2s^{-\frac{i\n^2}{2g}}\sim - g^{\frac{3}{2}}\sqrt{\frac{\p}{4}}(1+i) \frac{s^{-ig} }{ s\log ^{\frac{3}{2}}(s)}\,.
\end{split}
\end{align}
Further, at large coupling $\frac{\G\left((1\pm ig)/2\right)^2}{\G(\pm ig/2)^2}\sim\pm\frac{ig}{2}$, so that we obtain,
\begin{shaded}
\be
\mm_{(1)}^-\sim \left[\frac{4 c^4 (1+i)}{s\log^{\frac{3}{2}}(s)}\text{csch}(\pi  g)\sqrt{\frac{g}{\p}}\left\lbrace is^{ig}\,\frac{\G\left(\frac{3-t-ig}{2}\right)^2}{\G\left(\frac{3-t}{2}\right)^2}-s^{-ig}\,\frac{\G\left(\frac{3-t+ig}{2}\right)^2}{\G\left(\frac{3-t}{2}\right)^2}\right\rbrace\right]-(s\rightarrow -s).
\ee
\end{shaded}
\section{$2-$magnon correlator}\label{2mag}
For $2-$magnon case, we obtain the Regge Mellin amplitude from \eqref{melfn} by putting $\D_1=\D_2=\D_3=\D_4=2$,
	\begin{align} \label{2MagReg}
		\begin{split}
			\mm_{(2)}^{\pm}=\Bigg[&\frac{\pm1}{2\pi i}\oint dJ\frac{\pi}{\sin\pi J} \int_{-\infty}^{+\infty}d\n\left(\frac{s}{4}\right)^J \nu\,\sinh\pi\n ~\zeta_2({\D_i,t})\\
			&\times
			\frac{
				(J+1)  \Gamma (J-i \nu +2) \Gamma (J+i \nu +2) \Gamma \left(\frac{-J-t-i \nu +2}{2}\right) \Gamma \left(\frac{-J-t+i \nu +2}{2}\right)
			}
			{
				2\pi ^6 \Gamma \left(\frac{J-i \nu +2}{2}\right)^2 \Gamma \left(\frac{J+i \nu +2}{2}\right)^2
			}\frac{E_{2+i\n,J}^{(2)}}{1-\chi_2 E_{2+i\n,J}^{(2)}}\Bigg]\pm(s\rightarrow -s), 
		\end{split}
	\end{align}
\noindent
where from \eqref{eigs},
\be\label{2mageigen} 
E^{(2)}_{\D,J}=\frac{\psi_1(\frac{J-\D+4}{4})-\psi_1(\frac{J-\D+6}{4})-\psi_1(\frac{J+\D}{4})+\psi_1(\frac{J+\D+2}{4})}{ (4\pi)^4(J+1)(\D-2)}\,, \quad \chi_2=256\pi^4\xi^4\,,
\ee 
and $$\zeta_2({\D_i,t})=\frac{1}{\G(2-\frac{t}{2})^2}.$$
The Regge trajectories are then given by  poles of
\be\label{spec}
\frac{1}{\left(E_{2+i\n,J}^{(2)}\right)^{-1}-\chi_2 }\,.
\ee
Solving for the Regge trajectories for general coupling is complicated. However in the perturbative regime, analytical solution is tractable. We consider separately, weak ($\xi\rightarrow 0$) and strong coupling ($\xi\rightarrow \infty$) regimes.
\subsection{Weak Coupling}
For the weak coupling regime ($\xi\ll 1$), we can either have $\nu\gg O(\xi)$ and $\nu\sim O(\xi)$ which leads to two completely different perturbative solutions for the Regge poles. In the next two subsections, we will consider each of these sub-regimes in the weak coupling limit. We will elaborate on different solutions for the Regge poles and the schematics of the $\n$ integral briefly in the following.
\subsubsection{$\nu\gg O(\xi)$}
$J(\n)$ obtained from \eqref{spec} have four distinct sets of infinite trajectories. These are,
\be\label{pert0}
J_n(\pm\n)=\begin{cases} \pm i\n-2-4n+\sum_{k\geq1}\xi^{2k}\a_{n,k}(\pm \n),\\ \pm i\n-4-4n+\sum_{k\geq1}\xi^{2k}\b_{n,k}(\pm \n).\end{cases}\,
\ee
First few solutions for the two cases are,
\begin{align}\label{2magReggeweak}
\begin{split}
\a_{n,1}(\n)=&\pm \frac{4e^{i\pi/4}}{\sqrt{\n(1+4n-i\n)}}\,,\ \a_{n,2}(\n)=\frac{\a_{n,1}^2}{2(1+4n-i\n)}\,,\\
\a_{n,3}(\n)=&\frac{\a_{n,1}^3}{32(1+4n-i\n)^2}(20+(1+4n-i\n)^2(2\zeta_2-\psi_1(1/2-n)-\psi_1(1+n)\\
&-\psi_1(i\n/2-n)+\psi_1(i\n/2+1/2-n)))\,,\\
\b_{n,1}(\n)=&\pm \frac{4e^{-i\pi/4}}{\sqrt{\n(3+4n-i\n)}}\,,\ \b_{n,2}(\n)=\frac{\b_{n,1}^2}{2(3+4n-i\n)}\,,\\
\b_{n,3}(\n)=&\frac{\b_{n,1}^3}{32(3+4n-i\n)^2}(20+(3+4n-i\n)^2(2\zeta_2-\psi_1(-1/2-n)-\psi_1(1+n)\\
&-\psi_1(i\n/2-n)+\psi_1(i\n/2-1/2-n)))\,.
\end{split}
\end{align}
Observe that the four sets are effectively divided into two families which are related to each other by $\n\to-\n$.  This symmetry is due to the \enquote{shadow symmetry}(symmetry under $\D\to d-\D$) of \eqref{2mageigen}.
\paragraph{} Now while evaluating the $J$-integral in \eqref{2MagReg}, we make a transformation of the variable from $J$ to $\a_{n,1}$ or $\b_{n,1}$ depending on which family we are looking at. Consequently, our spectral functions \eqref{spec} takes the form, 
\be
\frac{E_{2+i\n,J(\pm \n)}^{(2)}}{1-\chi E_{2+i\n,J(\pm\n)}^{(2)}}=\begin{cases}\frac{i}{16\p^4\xi^4((1+4n\mp i\n)\n \a_{n,1}^2-16i)},\\ -\frac{i}{16\p^4\xi^4((3+4n\mp i\n)\n\b_{n,1}^2+16i)}.\end{cases}\,
\ee
with poles at $\a_{n,1}$ and $\b_{n,1}$ respectively. Note that the poles of $\alpha_{n,1}$ and $\beta_{n,1}$ lead to the perturbative solution. The corresponding Jacobian of transformation is,
\be
\Theta(\n)=\bigg|\frac{\pd J_n(\n)}{\pd\g_{n,1}}\bigg|\,, \ \ \text{where}\ \ \g_{n,1}=\a_{n,1}\ \text{or}\ \b_{n,1}\,.
\ee
Explicitly, the Jacobian of transformation is given by,
\be
\Theta(\n)=\xi^2\bigg(1+\xi^2\frac{\pd\g_{n,2}}{\pd\g_{n,1}}+\xi^4\frac{\pd\g_{n,3}}{\pd\g_{n,1}}+\dots\bigg)\,.
\ee
Note that if $|\n|\sim O(1)$, then the perturbative expansion in \eqref{pert0} converges for $0<\xi<1$. However, for $|\n|<1$, we expect a new perturbativ expansion even at weak coupling regardless of the explicit dependence of $\nu$ on the coupling. In the next section, we will derive the perturbative expansion when $|\n|<1$. 
\subsubsection{$\n\sim O(\xi^4)$}
Since the perturbation expansion breaks down when $\n\sim O(\xi^4)$ \footnote{To see this, note that when $\n\sim O(\xi^4)$, the subleading terms in \eqref{2magReggeweak} become comparable to the leading terms.} we consider the case when $\n= x\xi^4$ with $|x|\leq1$. For this regime, the following ansatz $J=\sum_n a_n  \xi^{\frac{4n}{3}}$\footnote{The choice for such an ansatz is motivated by the obeservation that at $\n=0$, the spectral function \eqref{2mageigen} admits such a perturbation expansion for the poles of $J(\n)$.}. gives, for the leading Regge trajectory, 
\begin{align}
\begin{split}
J(\n)=&-2+ \alpha _1 \xi ^{4/3}+\frac{1}{3} \alpha _1^2 \xi ^{8/3}+\frac{1}{3} \alpha _1^3 \xi ^4+\frac{\alpha _1^4 \xi ^{16/3} (1120-81 \zeta (3))}{2592}\\
&+\xi ^{20/3} \left(\alpha _1^5 \left(-\frac{5 \zeta (3)}{96}+\frac{7 \pi ^4}{11520}+\frac{154}{243}\right)-\frac{2 x^2}{3 \alpha _1}\right)+\cdots\,,
\end{split}
\end{align}
for which,
 
	\be \label{nuxi}
	\frac{E_{2+i\n,J}^{(2)}}{1-\chi E_{2+i\n,J}^{(2)}}=-\frac{1}{4 \left(\pi ^4 \left(\alpha _1^3+64\right)\right) \xi ^4}\,,
	\ee
and the Jacobian of transformation becomes,
\be \label{specd}
\begin{split}
	\Theta(\n)&=\xi ^{4/3}+\frac{2}{3} \alpha _1 \xi ^{8/3}+\alpha _1^2 \xi ^4+\xi ^{20/3} \left(\frac{2 x^2}{3 \alpha _1^2}+5 \alpha _1^4 \left(-\frac{5 \zeta (3)}{96}+\frac{7 \pi ^4}{11520}+\frac{154}{243}\right)\right)\nonumber\\
	&+\frac{1}{648} \alpha _1^3 \xi ^{16/3} (1120-81 \zeta (3))+\cdots
\,.
\end{split}
	\ee
The existence of two different perturbative solutions for the Regge spin for different regions of $\nu$ can be intuitively related to the level crossing phenomenon \cite{Korchemsky:2015cyx}.
\subsubsection{Evaluating the integral}\label{inteva}
To sum up, for $0<\xi<1$, the following two solutions for the leading Regge poles are relevant for our analysis,
\be \label{2magpolestotal}
J(\n)=\begin{cases} i\n-2-\sum_{k\geq1}\xi^{2k}\g^2_{0,k}\,, , \ |\n|>\xi^4,\\ -2+ \alpha _1 \xi ^{4/3}+\frac{1}{3} \alpha _1^2 \xi ^{8/3}+\frac{1}{3} \alpha _1^3 \xi ^4+\dots\,,\ |\n|\leq \xi^4,\end{cases}
\ee
where $\g^2_{0,k}$ are $\alpha_{n,k}$ for $n=0$ (explicitly given in \eqref{2magReggeweak}). In the weak coupling limit, \eqref{2MagReg} becomes,
\be
\mm_{(2)}^+=\frac{\zeta_2(\D_i,t)}{2\pi i}\int_{-\infty}^\infty d\n \n\sinh\pi \n\oint dJ\frac{\p}{\sin \pi J}\mathfrak{M}(J,\n)\left(\frac{s}{4}\right)^J \frac{E^{(2)}_{2+i\n,J}}{1-\xi^4 E^{(2)}_{2+i\n,J}}\, + (s \rightarrow -s)\,,
\ee
where for brevity and convenience, we denote the measure, 
\be
\mathfrak{M}(J,\n)=\frac{(J+1)\G(J-i\n+2)\G(J+i\n+2)\G(\frac{2+i\n-J-t}{2})\G(\frac{2-i\n-J-t}{2})}{2\pi^6\G(\frac{J-i\n+2}{2})^2\G(\frac{J+i\n+2}{2})^2}\,.
\ee
We split up the $\n-$integral into sub-regimes where we will separately solve for the spectral function,
\begin{align}
\begin{split}
\mm_{(2)}^+=&\frac{\zeta_2(\D_i,t)}{2\pi i}\left(\int_{-\infty}^{-{\xi^4}}+\int_{\xi^4}^\infty\right) d\n \n\sinh\pi \n\oint dJ\frac{\p}{\sin \pi J}\mathfrak{M}(J,\n)\left(\frac{s}{4}\right)^J \left(\frac{E^{(2)}_{2+i\n,J}}{1-\xi^4 E^{(2)}_{2+i\n,J}}\right)_{|\n|>\xi^4}\\
&+\frac{1}{2\pi i}\int_{-{\xi^4}}^{\xi^4} d\n \n\sinh\pi \n\oint dJ\frac{\p}{\sin \pi J}\mathfrak{M}(J,\n)\left(\frac{s}{4}\right)^J \left(\frac{E^{(2)}_{2+i\n,J}}{1-\xi^4 E^{(2)}_{2+i\n,J}}\right)_{|\n|\leq \xi^4}+ (s \rightarrow -s)\,, \\
=&A_1+A_2\,,
\end{split}
\end{align}
where $A_1$ denotes the regime for $|\n|>{\xi^4}$ and $A_2$ denotes the regime for $|\n|\leq{\xi^4}$. The integral receives contributions to from both $A_1$ and $A_2$. We will start with writing the integral $A_2$ which is, 
\begin{align}  
\begin{split}
A_2&=\frac{\zeta_2(\D_i,t)}{2\pi i}\int_{-{\xi^4}}^{\xi^4} d\n \n\sinh\pi \n\oint dJ\frac{\p}{\sin \pi J}\mathfrak{M}(J,\n)\left(\frac{s}{4}\right)^J \left(\frac{E^{(2)}_{2+i\n,J}}{1-\xi^4 E^{(2)}_{2+i\n,J}}\right)_{|\n|\leq {\xi^4}}\,.
\end{split}
\end{align}
The Regge pole is given by, 
\begin{align}
	\begin{split}
	J(\n)=&-2+ \alpha _1 \xi ^{4/3}+\frac{1}{3} \alpha _1^2 \xi ^{8/3}+\frac{1}{3} \alpha _1^3 \xi ^4+\xi ^{20/3} \left(\alpha _1^5 \left(-\frac{5 \zeta (3)}{96}+\frac{7 \pi ^4}{11520}+\frac{154}{243}\right)-\frac{2 x^2}{3 \alpha _1}\right)+\dots
	\end{split}
	\end{align}
Transforming $dJ\rightarrow d\a_1$, along with \eqref{nuxi} and \eqref{specd}, we can write, 
\be
A_2=\frac{\zeta_2(\D_i,t)}{2\pi i}\int_{-\xi^4}^{\xi^4} d\n \n\sinh\pi\n\oint d\a_1\frac{\p}{\sin\pi J(\a_1)}\Theta(\a_1,\n)\mathfrak{M}(\a_1,\n)\left(\frac{s}{4}\right)^{J(\a_1)}\left( -\frac{1}{4 \left(\pi ^4 \left(\alpha _1^3+64\right)\right) \xi ^4}\right)\,,
\ee
In order to perfrom the integral we make the change of variables $\nu \rightarrow x \xi^4$,
\begin{align}\label{A2}
\begin{split}
A_2&=\left(-\frac{\xi ^{32/3}}{2304 \pi ^{9} L^2}+\frac{\xi ^{12}  \left(\log (L)-\psi ^{(0)}\left(2-\frac{t}{2}\right)\right)}{576 \pi ^{9} L^2 }\right.\\
&\left.-\frac{\xi ^{40/3}}{5184 \pi ^{9} L^2}  \left(-12 (3 \log (L)+1) \psi ^{(0)}\left(2-\frac{t}{2}\right)+18 \log ^2(L)+12 \log (L) +18 \psi ^{(0)}\left(2-\frac{t}{2}\right)^2\right.\right.\\
&\left.\left. +9 \psi ^{(1)}\left(2-\frac{t}{2}\right)+9 \pi ^2+8\right)\right)
\end{split}
\end{align}
Now we turn to $A_1$. The detailed logic for the evaluation of this is given in appendix \ref{2magdetails}. We quote here the final result, 
\be 
A_1= -\frac{\xi ^{12}}{576\p^9L^2}\left[3\log L-18 \psi ^{(0)}\left(2-\frac{t}{2}\right)\right]+O(\x^{20}).
\ee 
The equality here is in the sense of asymptotic equivalence in the limit $L\to\infty$. Finally adding $A_1$ and $A_2$ we obtain, 
\begin{align}\label{m2weak}
\begin{split}
\mm_{(2)}^+&=L^{-2}
\left\lbrace
\left[
-\xi ^{32/3}\frac{1}{2304 \pi ^{9}}
+\xi ^{12}\frac{17 }{576 \pi ^{9}}\psi ^{(0)}\left(2-\frac{t}{2}\right)
+\frac{\x^{40/3}}{5184\p^9}
\left(12\psi ^{(0)}\left(2-\frac{t}{2}\right)-18\psi ^{(0)}\left(2-\frac{t}{2}\right)^2
\right.
\right.
\right.\\
&\left.
\left.
\hspace{2.3 cm}-9\psi ^{(1)}\left(2-\frac{t}{2}\right)-9\p^2-8\right)+\dots
\right]
+\log (L)\left[-\frac{\xi ^{12} }{288 \pi ^{9}}+\frac{\x^{40/3}}{216} +\dots \right]
\\
&\left.\hspace{7.8 cm}+\log^2L\left[-\frac{\x^{40/3}}{288\p^9}+\dots\right]+O(\log^3L)
\right\rbrace+~(s\to-s)
\end{split}
\end{align}
where the dots represent terms subleading in $\x$. 
\subsection{Strong Coupling}
We will now investigate the strong coupling regime $\x\gg1$. There are similar two regions of interest $\n\ll\x$ and $\n\sim O(\x)$. We will analyze these two cases separately below.
\subsubsection{$\n\ll O(\x)$}
For, $\n\ll O(\x)$, we consider an expansion around $\x\rightarrow \infty$ keeping $\nu$ fixed. For strong coupling $\xi\rightarrow \infty$, the denominator of  
\eqref{spec} can be written in a summation representation as,
\be 
\mathcal{E}_2=(4\pi)^4 E_{2+i\n,J}^{(2)}=\frac{1}{(J+1)} \sum_{n=0}^{\infty} \frac{(-1)^{n}(2 n+J+2)}{(i\n-2 n-J-2)^{2}(2+i\n+2 n+J)^{2}}.
\ee
Since $\mathcal{E}_2\sim 1/J^4$ for large $J$, we can make an ansatz,
\be 
J= a_0+ a_1\xi+\sum_{n=1}^{\infty}\frac{a_{-n}}{\xi^n}.
\ee    
Putting this ansatz in \eqref{spec}, we obtain that the solutions for $a_1$, 
\be \label{admissible}
a_1= \pm 2\sqrt[4]{2},\pm 2i\sqrt[4]{2}.
\ee 
We neglect $a_1=-2\sqrt[4]{2}$ since the exponent of $s$ for this root is extremely subleading compared to the others. The coefficients are obtained recursively as (we evaluate explicitly upto order $\xi^{-3}$),
\be \label{large0}
a_0= -1\,, \ a_{-1}=-\frac{3+\n^2}{2a_1}\,,\ a_{-2}=0\,,\ a_{-3}=\frac{87+18\n^2-\n^4}{8a_1^3}\,.
\ee
For each value of $a_1$, we obtain separate solutions for $J$. For example, $a_1=2\sqrt[4]{2}$ gives ,
\be \label{large1}
J= -1+\left[2\sqrt[4]{2}\xi -\frac{\nu ^2+3}{4 \sqrt[4]{2}\xi}+\frac{87+18 \nu ^2-\nu ^4}{64 \sqrt[4]{8}\xi^3}+\mathcal{O}\left(\frac{1}{\xi^4}\right)\right]\,,
\ee 
while $a_1=\pm i 2\sqrt[4]{2}$ gives,
\be \label{large2}
J=-1\pm i\left[2\sqrt[4]{2}\xi+\frac{\n^2+3}{4\sqrt[4]{2}\xi}+\frac{87+18 \nu ^2-\nu ^4}{64 \sqrt[4]{8}\xi^3}+\mathcal{O}\left(\frac{1}{\xi^4}\right)\right]\,.
\ee 
Note that the expansions above (including \eqref{large1}, \eqref{large2}) are valid for $|\n|<1$ but breaks down for $|\n|\sim O(\xi)$.
\subsubsection{$\n\sim O(\x)$}
For large $|\n|$ and large $\x$ the two are loosely related by $\n\sim O(\x)$\footnote{Precisely speaking $|\n|>1$ and $\x>1$ are distinct. }. In this case , wee will consider a different expansion for the Regge poles. It is a double expansion,
\be 
J\sim -1+g(\n,\x),
\ee 
with,
\begin{align} 
\begin{split}
&g(\n,\x)=a_{1}+\frac{3\n^2-3 a_1^2}{2a_1(a_1^2+\nu ^2)}+\frac{33 a_1^2 \nu ^4-311 a_1^4 \nu ^2+87 a_1^6-9 \nu ^6}{8 a_1^3 \left(a_1^2+\nu ^2\right){}^3}+O\left(\frac{1}{a_1^{5}(a_1^2+\n^2)^{5}}\right)
\end{split}
\end{align}
The solutions for $a_{1}$ are,
\be 
a_{1}=\, \pm i\sqrt{4\sqrt{2}\x^2+\n^2},\,\pm \sqrt{4\sqrt{2}\x^2-\n^2}
\ee 
\subsubsection{Evaluation of the Mellin amplitude}
For strong coupling, one can take $|\xi|\gg1$ for all practical purposes. Since the $\nu-$integral (from $(-\infty, \infty)$), has distinct regions for $\nu\sim\xi$, $\nu\ll\xi$ and $\nu>\xi$,
\be
M_{(2)}^\pm=\left[\frac{\pm1}{2\pi i}\int_{-\infty}^\infty d\n \n\sinh\pi \n\oint dJ\frac{\p}{\sin(\p J)}\mathfrak{M}(J,\n;t)\left(\frac{s}{4}\right)^J \frac{E^{(2)}_{2+i\n,J}}{1-\xi^4 E^{(2)}_{2+i\n,J}}\right]\pm(s\to-s)\,,
\ee
with,
\be
\mathfrak{M}(J,\n;t)=\frac{(J+1)\G(J-i\n+2)\G(J+i\n+2)\G(\frac{2+i\n-J-t}{2})\G(\frac{2-i\n-J-t}{2})}{2\pi^6\G(\frac{J-i\n+2}{2})^2\G(\frac{J+i\n+2}{2})^2}\zeta_2(\D_i,t)\,,
\ee
can be subdivided according to these regimes. For $\n\ll\xi$, \eqref{large0} becomes,
\be \label{Jstrong}
J(\n)=-1+a_1\xi-\frac{3+\n^2}{2a_1\xi}+\frac{87+18\n^2-\n^4}{8a_1^3\xi^3}+\dots\,.
\ee
with the Jacobian of transformation,
\be
\Theta=\frac{\pd J(\n)}{\pd a_1}=\xi\left(1+\frac{3+\n^2}{2a_1^2\xi^2}-\frac{3(87+18\n^2-\n^4)}{8a_1^4\xi^4}+\dots\right)\,.
\ee
Now define $J_R=-1+a_1\xi$, so that an expansion in $1/\x$ about $J_R$ in the limit $\x\to\infty$ gives, 
\begin{align}
\begin{split}
\frac{\mathfrak{M}(J_R,\n)}{\zeta(\D_i,t)}=&\frac{\csc (\pi J_R) \Gamma (J_R-i \nu +2) \Gamma (J_R+i \nu +2) \Gamma \left(\frac{-J_R-t-i \nu +2}{2}\right) \Gamma \left(\frac{-J_R-t+i \nu +2}{2}\right)}{2 \pi ^6 \Gamma \left(\frac{J_R-i \nu +2}{2}\right)^2 \Gamma \left(\frac{J_R+i \nu +2}{2}\right)^2}(J_R+1)\\
&\times\Bigg(1 +\frac{3+\n^2}{2a_1\xi}\bigg[(H_{\frac{J_R+i\n}{2}}+H_{\frac{J_R-i\n}{2}}-H_{1+J_R+i\n}\\ &-H_{1+J_R-i\n}+\frac{1}{2}H_{\frac{-J_R-t+i\n}{2}}+\frac{1}{2}H_{\frac{-J_R-t-i\n}{2}}-\g+\pi\cot(\pi J_R))-\frac{1}{J_R+1}\bigg]+\dots\Bigg)\,.
\end{split}
\end{align}
Thus, to the leading order,
\begin{align}
\begin{split} 
\frac{\mathfrak{M}(J_R,\n)}{\zeta(\D_i,t)}\approx \frac{\csc (\pi J_R) \Gamma (J_R-i \nu +2) \Gamma (J_R+i \nu +2) \Gamma \left(\frac{-J_R-t-i \nu +2}{2}\right) \Gamma \left(\frac{-J_R-t+i \nu +2}{2}\right)}{2 \pi ^6 \Gamma \left(\frac{J_R-i \nu +2}{2}\right)^2 \Gamma \left(\frac{J_R+i \nu +2}{2}\right)^2}(J_R+1)\,.
\end{split}
\end{align} 
However, as for the exponent of $(s/4)$ we will consider at least upto order $1/\x$. This is because since this is in exponent, the variation over $\x$ is stronger than that in $\mathfrak{M}$.
\paragraph{}Further, 
\be\label{residue}
\frac{E^{(2)}_{2+i\n,J}}{1-\xi^4 E^{(2)}_{2+i\n,J}}=\frac{1}{8\p^4\xi^4(a_1^4-32)}\,,
\ee
with poles at $a_1= 2 \sqrt[4]{2}\,, \pm i2  \sqrt[4]{2}$ where as argued below \eqref{admissible}, we are not considering the negative real pole. We would like to emphasize here that given \eqref{Jstrong}, \eqref{residue} is in fact exact coupling. We can further this argument by expanding around $\n=0$ (assuming that the $\n-$integral is peaked around the origin\footnote{This is true for all practical purposes.} and neglecting the effect of the poles)\footnotemark,
\begin{align}
\begin{split}
\frac{M_{(2)}}{\zeta_2(\D_i,t)}&\sim\frac{\p}{2i\pi  \xi^2}\int_{-\infty}^\infty d\n \n^2 \left(\frac{s}{4}\right)^{-\frac{3+\n^2}{2a_1\xi}+\dots}\oint \frac{da_1 a_1}{8\p^4(a_1^4-32)}~\frac{\G(2+J_R)^2\G(1-\frac{J_R+t}{2})^2}{2\p^5\sin(\p J_R)\G(1+J_R/2)^4}\left(\frac{s}{4}\right)^{J_R},\\
&\sim-\frac{1}{16 \sqrt{2}i \pi ^{9}}\frac{1}{s\log^{\frac{3}{2}}s}\sqrt{\frac{\x}{\p}}\left[\oint da_1 \frac{a_1^{7/2}4^{a_1\x} }{a_1^4-32}\csc(\p a_1\x)\left(\frac{s}{4}\right)^{a_1\x-\frac{3}{2a_1\xi}}\G\left(\frac{3-t}{2}-\frac{a_1\x}{2}\right)^2\right].
\end{split}
\end{align}
\footnotetext{Note that, here we have written the expression excluding all the overall sign factors and $s\to-s$ factor.} This $\log s$ dependence is crucial and matches with the strong coupling analysis of $0,1-$magnon cases. The rest of the power law analysis can be obtained in a straightforward manner by simply picking out the residues of the $a_1-$integral.
Observe that the dominant contribution will be given by $a_1=2\sqrt[4]{2}$, while the others give a phase. Therefore in $\x\to \infty$ limit,
\begin{shaded} \label{2magstrong}
	\be
	M_{(2)}^{+}\sim \left[-\frac{1}{16\ 2^{7/8}\p^{9}}\sqrt{\frac{\x}{\p}}\frac{s^{2 \sqrt[4]{2} \xi  }}{s\log^{\frac{3}{2}}s}\csc \left(2 \sqrt[4]{2} \pi  \xi \right)\frac{\G\left(\frac{3-t}{2}-\sqrt[4]{2}\x\right)^2}{\G\left(\frac{4-t}{2}\right)^2}\right]+ (s\to-s).
	\ee
\end{shaded}
\section{Comparison among $0, 1-$magnon Regge trajectories}
\paragraph{} 
	\begin{figure}[!htb] 
	\centering
	\vspace{0.5cm}
	\begin{subfigure}{.5\textwidth}
		\centering
		\includegraphics[width=.9\linewidth]{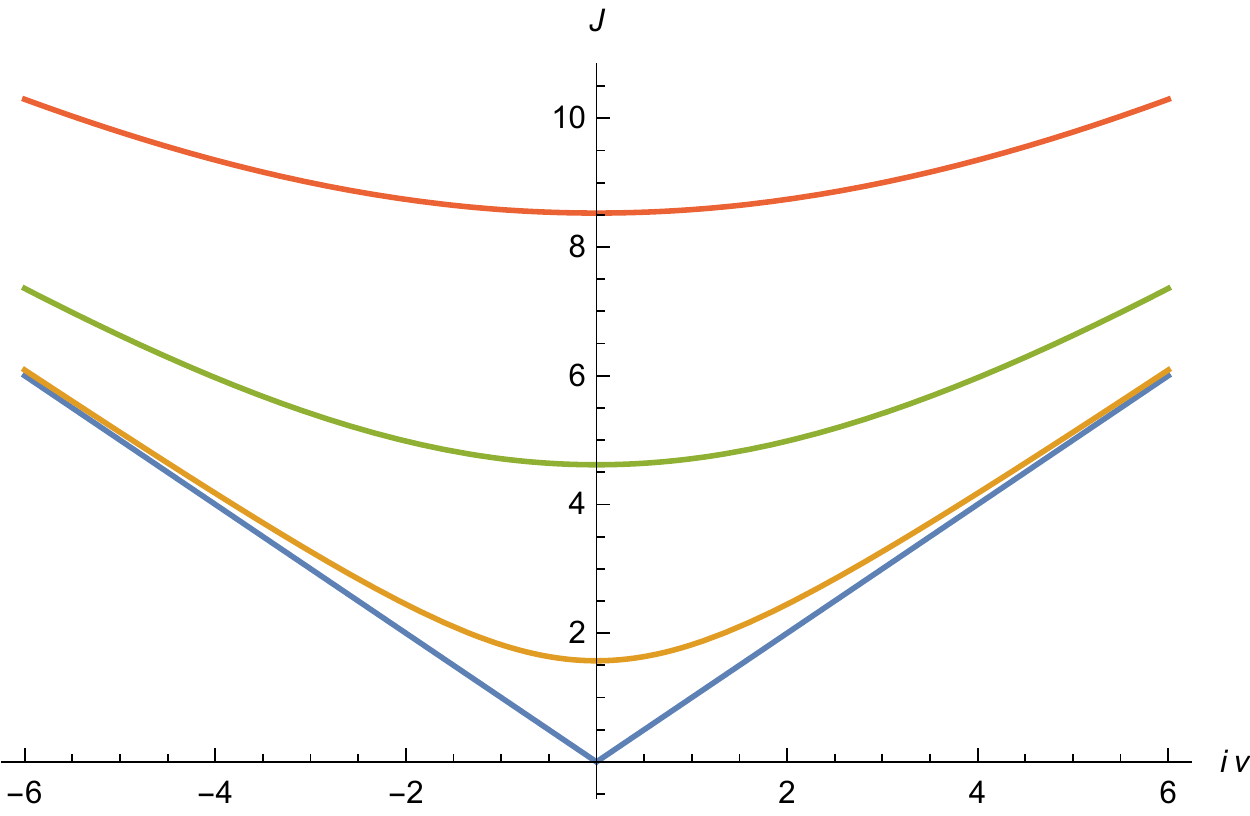}
		\caption{ 0-magnon }
		\label{fig:sub1}
	\end{subfigure}%
	\begin{subfigure}{.5\textwidth}
		\centering
		\includegraphics[width=.9\linewidth]{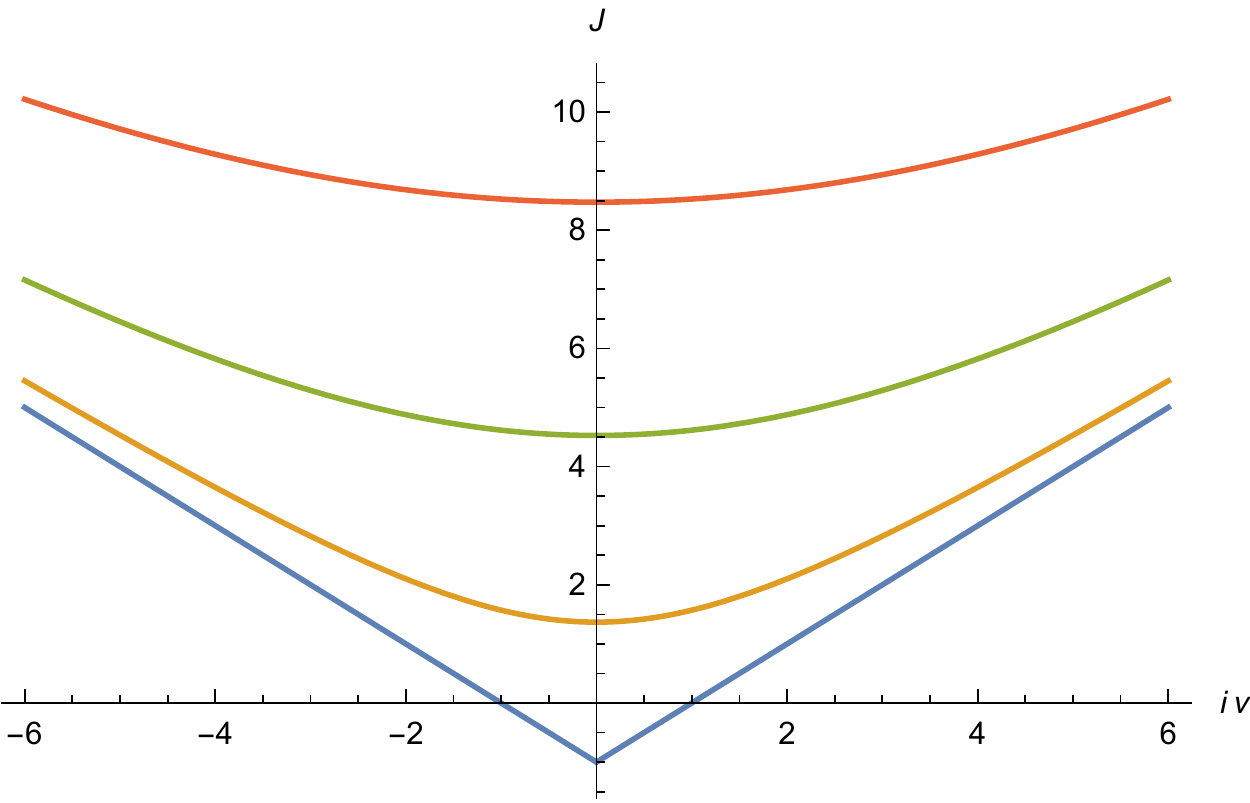}
		\caption{ 1-magnon }
		\label{fig:sub2}
	\end{subfigure}
	\caption{Leading Regge trajectories for 0 and 1 magnon correlators in the weak coupling. The chosen values of couplings are $\x=0$(blue), $\x=0.03$(orrange), $\x=0.07$(green), $\x=0.12$(red).}
	\label{fig:test}
\end{figure}
 Before concluding, we will compare between the leading Regge trajectories for the $0,1-$magnon correlators for the fishnet theory. The leading Regge trajectories, for the magnon correlators are characterized by,
\begin{itemize}
	\item $0-$magnon:
	\be
	j_0(\n)=-1+\sqrt{1-\n^2+ 2\sqrt{f^4-\n^2}}\,.
	\ee 
	\item $1-$magnon:
	\be
	j_1(\n)= -1+ \sqrt{g^2-\n^2}\,. 
	\ee
	The growth of Regge spin  for both $0$ and $1$ magnon correlators with $\nu$ has been plotted in figure \ref{fig:test}. Note that the plots are given in terms of the reduced coupling $\x$ which are related to the relevant couplings for $0-$ and $1-$magnon by $f=4\sqrt{2}c\p^2\x$ and $g=8\p^2c\x$ respectively. Observe the obvious shift in the intercept which is clear from the weak coupling expressions of the respective Regge trajectories.  	
\end{itemize}
For the $2-$magnon case, the solutions are known analytically only perturbatively. The analysis we have performed, extracts the analytical behaviour for the leading Regge poles in the weak and strong coupling which excludes the need of a graphical description. It would however be interesting to see graphically (numerically), how to patch up the solutions in various regimes.

\section{Discussions}
\paragraph{} We present the salient observations of our exercise in the following.
\begin{itemize}
	\item We have considered the Regge limit of the $0,1,2-$magnon correlators in the four dimensional conformal fishnet theory.  The techniques of Conformal Regge Theory in Mellin space as expounded in \cite{Costa2012a} has been deployed along the lines of \cite{Korchemsky2018} in order to derive the weak coupling expansion for the fishnet correlators. For $0,1-$magnon correlators, we find exact Regge trajectories and compute the Regge limit of the Mellin amplitude in the weak coupling. For the strong coupling limit we do an order of magnitude computation with regards to the leading behavior. For 0-magnon correlator the results we have obtained, match with the analysis of \cite{Korchemsky2018} in both regimes of coupling subject to the conditions described in subsubsection \ref{cwer0m}.  
\item For the $2-$magnon case, solving the spectral function for any finite value of the coupling seems a formidable task in contrast with the $0,1-$magnon correlators. However, a systematic expansion in the weak/strong coupling limit is still possible. We have analyzed the weak coupling limit in detail while for the strong coupling we have naively compared the leading power law singularity in the Regge limit (along the lines of \cite{Korchemsky2018}).  
\item  In comparison with \cite{Korchemsky2018}, we would like to point out one subtle difference. \cite{Korchemsky2018} used the LSZ-type prescription to analyze the on-shell scattering amplitude. For the $0-$magnon case, every exchange including the external operators are on-shell. For the $1,2-$magnon case, some or all of the external operators are off-shell as explained in the introduction. Though we have analyzed the Regge limit of the correlators themselves using the techniques of \cite{Costa2012a} thereby bypassing the LSZ-type analysis in \cite{Korchemsky2018}, it is worth of investigating whether we can devise systematic perturbative methods in terms of Feynman Diagrams for the $1,2-$magnon case.
\item Our analysis can be extended straightforwardly to various cases of conformal fishnet theories\footnote{We thank Vladimir Kazakov and Antonio Pittelli for suggesting these.}. These include fishnet theories in general dimensions \cite{Kazakov:2018qez}, chiral fishnet theories \cite{Kazakov:2018gcy}, fishnet theory obtained from four-dimensional $\mathcal{N}=2$ quiver gauge theories \cite{Pittelli:2019ceq}. Similar analysis can be undertaken for the double scaling limit of $\g$-twisted ABJM theories considered in \cite{Caetano:2016ydc}. 
\item Another possible direction is to compare the strong coupling results (the order to magnitude of the leading term) with the holographic counterpart {\it i.e.} the quantum holographic fishchain model recently discussed in \cite{Gromov:2019bsj,Gromov:2019aku}.
\item It would be nice to correlate these Regge trajectories of the \enquote{n}-magnon correlators to known results for ${\cal N}=4$ SYM \footnote{We thank Nikolay Gromov for suggesting this to us.}.
\end{itemize}
\section{Acknowledgements}
We thank Gregory Korchemsky for useful discussions during initial stages of the work and very important comments and questions regarding the draft. We also thank Nikolay Gromov, Aninda Sinha and Vladimir Kazakov for useful comments on the draft. SDC thanks Abhijit Gadde for discussions. KS thanks Simon Caron-Huot for discussions during the initial stages of the work. KS is partially supported by World Premier Research Center Initiative (WPI) initiative, MEXT Japan at Kavli IPMU, World Premier Research Center Initiative (WPI) the University of Tokyo.

\appendix
\section{Details of Pole  Analysis}\label{polestructure}
\paragraph{ }In this Appendix, we review the contour manipulation that the authors of \cite{Korchemsky2018} use to compute the Regge limit at weak coupling and extend it to our analysis of Mellin amplitudes. 
\subsection{Details of $0$-Magnon Analysis }\label{poleanalysis0mag}
\paragraph{} In this subsection we will explain the details of how do we reach the equation \eqref{zero1}. We follow essentially \cite{Korchemsky2018} and review the method for our case. We start with 
\be \label{zero2}
\mathcal{A}=\int_{-\infty}^{\infty} d\n \left[\left(\frac{s}{4}\right)^{J_2^+}F(\n,J_2^+)+\left(\frac{s}{4}\right)^{J_2^-}F(\n,J_2^+)\right]
\ee 
with
\be 
J_{2}^{\pm}=-1+\sqrt{1-\n^2\pm 2\sqrt{f^4-\n^2}}
\ee 
and ,
\be 
F(\n,J)=\frac{ \nu   \sinh (\pi  \nu ) \Gamma (J-i \nu +2) \Gamma (J+i \nu +2) \Gamma \left(\frac{-J-t-i \nu +2}{2}\right) \Gamma \left(\frac{-J-t+i \nu +2}{2}\right)}{\sin (\pi  J) \left(J (J+2)+\nu ^2\right) \Gamma^2 \left(\frac{J-i \nu +2}{2}\right) \Gamma^2 \left(\frac{J+i \nu +2}{2}\right)}
\ee 
First for brevity we define,
\be 
\F_{\pm}(\n)=\left(\frac{s}{4}\right)^{J_2^{\pm}(\n)}F(\n,J_2^{\pm}(\n))
\ee 
We split the integration region in \eqref{zero2} as following,
\begin{align}\label{A.5}
\begin{split}
\mathcal{A}=&\int_{-f^2}^{f^2}d\n\left[\F_{+}(\n)+\F_{-}(\n)\right]+\left(\int_{-\infty}^{-f^2}d\n\F_{+}(\n)+\int_{f^2}^{-\infty}d\n\F_{-}(\n)\right)\\
&\hspace{5 cm}+\left(\int_{-\infty}^{-f^2}d\n\F_{-}(\n)+\int_{f^2}^{-\infty}d\n\F_{+}(\n)\right)
\end{split}
\end{align} 
The key step is to show that at large $s$,
\begin{eqnarray}\label{A.6}
\begin{split}
\int_{-\infty}^{-f^2}d\n\F_{+}(\n)+\int_{f^2}^{\infty}d\n\F_{-}(\n)&=&-\int_{-f^2}^{f^2}d\n\,\F_{-}(\n)+O\left(\frac{1}{s}\right)\\
\int_{-\infty}^{-f^2}d\n\F_{-}(\n)+\int_{f^2}^{-\infty}d\n\F_{+}(\n)&=&-\int_{-f^2}^{f^2}d\n\,\F_{-}(\n)+O\left(\frac{1}{s}\right)
\end{split}
\end{eqnarray}
where the second relation follows from first one upon replacing $\n\rightarrow -\n$ and taking into account that $\F_{\pm}(-\n)=\F_{\pm}(\n)$. If we now substitute \eqref{A.6} into \eqref{A.5} then we obtain,
\begin{align}\label{append1}
\begin{split}
\mathcal{A}&=\int_{-f^2}^{f^2}d\n\, \left[\F_{+}(\n)+\F_{-}(\n)\right]-2\int_{-f^2}^{f^2}d\n\F_{-}(\n)+O\left(\frac{1}{s}\right)\\
&=\int_{-f^2}^{f^2}d\n\, \left[\F_{+}(\n)-\F_{-}(\n)\right]+O\left(\frac{1}{s}\right)
\end{split}
\end{align}
To prove \eqref{A.6} first we introduce the change variable $\n^2-\f^4=x^2$ so that,
\be 
J_{2}^{\pm}=-1+\sqrt{1-f^4-x^2\pm 2ix}
\ee  
With this change of variable  \eqref{A.6} becomes,
\be 
\int_{-\infty}^{-f^2}d\n\F_{+}(\n)+\int_{f^2}^{-\infty}d\n\F_{-}(\n)=2\, \text{Re} \int_{0}^{\infty}\frac{x\,dx}{\sqrt{x^2+f^4}}\F_{-}(\sqrt{x^2+f^4})
\ee 
Here we took into account that $J_{2}^+$ and $J_{2}^-$ are conjugate to each other for real $x$ such that $1-f^4-x^2>0$. \\
In a similar fashion, the integral on the right-hand side in the first line of (\ref{A.6}) we find upon changing the variable $\n^2-f^4=-x^2$,
\be 
-\int_{-f^2}^{f^2}d\n\F_{-}(\n)=-2\int_{0}^{f^2}d\n\F_{-}(\n)=-2\int_{0}^{f^2}\frac{dx\, x}{\sqrt{f^4-x^2}} \F_-({\sqrt{f^4-x^2}})
\ee 
Now to match (A.10) into (A.9) we will rotate the integration contour in the integral (A.9). Before that we need to understand the contour prescription of the integral in (A.9) a bit. To get a hold of in which way we need to close the contour we observe that in the large $x$ limit we have,
\be 
\F_{-}(\sqrt{x^2+f^4})\sim \left(s/4\right)^{-ix},\qquad x\rightarrow \infty
\ee 
This suggests that we would like to close the $x$-contour in the lower half of complex $x-$plane in (A.9) . The contour that we will use is as below,
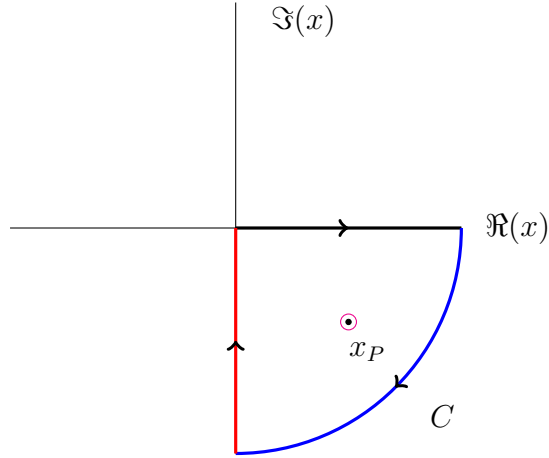
\begin{figure}[h]
	\centering
	\begin{tikzpicture}[scale=0.5]
	\draw[-] (-6,0)--(6,0);
	\draw[-] (0,-6)--(0,6);
	\draw (7.5,0) node {$\Re(x)$};
	\draw (1.8,5.5) node {$\Im(x)$};
	\draw (5.5,-5) node {$C$};
	\draw
	[
	very thick,
	decoration={markings, mark=at position 0.5 with {\arrow[black,line width=0.5mm]{>}}},
	postaction={decorate}
	]
	(0,0)--(6,0);
	\draw
	[
	very thick,
	color=blue,
	decoration={markings, mark=at position 0.5 with {\arrow[black,line width=0.5mm]{>}}},
	postaction={decorate}
	]
	(6,0) arc[start angle=0, end angle=-90, radius=6cm];
	\draw
	[
	very thick,
	color=red,
	decoration={markings, mark=at position 0.50 with {\arrow[black,line width=0.5mm]{>}}},
	postaction={decorate}
	]
	(0,-6)--(0,0);
	\filldraw  (3,-2.5)circle[radius=2pt];
	\draw[color=magenta] (3,-2.5)circle[radius=6pt];
	\draw (3.5,-3.3) node {$x_P$};
	\end{tikzpicture}
	\caption{Contour Prescription for (A.9)}
\end{figure}\\
Now, referred to the above contour prescription, we have
\begin{align}
\begin{split}
&\int_{0}^{\infty}\frac{x\,dx}{\sqrt{x^2+f^4}}\F_{-}(\sqrt{x^2+f^4})\\
&=\int_{0}^{-i \infty}\frac{x\,dx}{\sqrt{x^2+f^4}}\F_{-}(\sqrt{x^2+f^4})-2\p i\sum_{x_P}\text{Res.}\left[\frac{x}{\sqrt{x^2+f^4}}\F_{-}(\sqrt{x^2+f^4})\right]
\end{split}
\end{align}
where $\lbrace x_P\rbrace$ are the  poles of $\F_{-}(\sqrt{x^2+f^4})$ in $x$.\\
Observe that we have closed the contour in the lower half plane to ensure that the integral over $C$, which is a semi-circular arc of infinite radius, vanish. Also note that the residue sum comes with an overall negative sign because we have closed the contour in the clockwise sense.\\
Now it is very clear from the above representation that only those poles which lie in the lower half-plane, as shown in the figure, i.e, the poles with negative imaginary parts can contribute to the residue sum i.e, the poles that contribute have the generic structure,
\be 
x_P=\Re(x_P)-i\bar{\Im}(x_P),\qquad \bar{\Im}(x_P)>0
\ee
Next we observe that at these poles, the residue  give negative exponents of $s$ \textit{at weak coupling}. \textit{We would like to point this out specifically that this is only the case unanimously in the weak coupling regime, around $f\rightarrow 0$. At strong coupling things are  not so. Therefore the following reasoning that we are going to present will go through in the weak coupling limit}\footnote{But we took advantage of \eqref{zero1} in weak coupling anyway. So we are not bothered here about strong coupling!}.\\
Now with the above in place , these contributions are exponentially suppressed compared to the line integral in the Regge limit $s\rightarrow\infty$ i.e, the Regge limit. These are the $O(s^{-1})$ terms we wrote explicitly in \eqref{append1} and we are going to neglect these terms in Regge limit. Hence forth while writing we will not write these pole contributions, if any, explicitly and \textit{any equality will be understood modulo  contributions coming from these poles}. \\
Now we introduce the \enquote{Wick Rotation} $x=-ix_E$ and finally obtain from (A.9),
\be 
-2\text{Re}\int_{0}^{\infty}\frac{dx_E\,x_E}{\sqrt{-x_E^2+f^4}}\F_{-}\left(\sqrt{-x_E^2+f^4}\right)
\ee
The
integrand has two square-root branch cuts $[-\infty,-f^2)$ and $[f^2,\infty)$ and deforming the contour we should not cross the cut.\\
Next we split up (A.14),
\begin{align}
\begin{split}
-2\text{Re}\int_{0}^{\infty}\frac{dx_E\,x_E}{\sqrt{-x_E^2+f^4}}\F_{-}\left(\sqrt{-x_E^2+f^4}\right)&= -2\text{Re}\int_{0}^{f^2}\frac{dx_E\,x_E}{\sqrt{-x_E^2+f^4}}\F_{-}\left(\sqrt{-x_E^2+f^4}\right)\\
&-2\text{Re}\int_{f^2}^{\infty}\frac{dx_E\,x_E}{\sqrt{-x_E^2+f^4}}\F_{-}\left(\sqrt{-x_E^2+f^4}\right)
\end{split}
\end{align}
To proceed further, we use a crucial observation about the \enquote{physical spectrum of $t$}. The vital information is that the physical spectrum for $t$ consists of \textit{real values only}. And henceforth we will base our analysis on the physical spectrum of $t$. With this piece of information we observe that the collections of Gamma functions in $\F_-(x_E)$ come in the combination,
\be 
\G(p+iq)\G(p-iq),\qquad p,q\in\mathbb{R}
\ee 
with suitable values for $p,q$ \\
Since we have (this can be proved for instance using the Euler integral representation of Gamma function)
\be 
\G(z^*)=\G(z)^*
\ee 
so that 
\be 
\G(p+iq)\G(p-iq)=|\G(p+iq)|^2\,\in\mathbb{R}
\ee  \\
Hence, $\F_-(x_E)$ is real over the entire interval $y\in [0,\infty)$. However the factor 
\be 
\frac{x_E}{\sqrt{-x_E^2+f^2}}
\ee  is purely real for $y_E\in [0,f^2]$ but is purely imaginary for $y_E\in [f^2,\infty)$. Thus the piece of integral in (A.15) over the interval $ [f^2,\infty)$ vanishes identically and we have the left-hand side of (A.9) and (A.10) coincide upto corrections that vanish in $s\rightarrow \infty$.\\
Hence we have the desired relation \eqref{append1}.
\subsection{Details of $1$-Magnon Analysis }\label{poleanalysis1mag}
In this subsection we will deliver the details of the manipulation leading to the equation \eqref{1even}.We start with looking  into the following integral,
\begin{eqnarray}
I_1 =\int_{-\infty}^{-g} d\n F(J_e^+)s^{J_e^+} + \int_{g}^{\infty} d\n F(J_e^-)s^{J_e^-}
\end{eqnarray}
Because the integrand is even under $\left(\nu \rightarrow -\nu \right)$, we have
\begin{eqnarray}
I_1 =\int_{g}^{\infty} d\n F(J_e^+)s^{J_e^+} + \int_{g}^{\infty} d\n F(J_e^-)s^{J_e^-} 
\end{eqnarray}
Under the transformation of variable $\nu^2 -g^2= y^2$,
\begin{eqnarray}
I_1 &=& \int_{0}^{\infty} \frac{ydy}{\sqrt{y^2+g^2}}\left[  F(J_e^+(y))s^{J_e^+(y)} +  F(J_e^-(y))s^{J_e^-(y)}\right] \nonumber\\
&=&2 \textrm{Re}\int_{0}^{\infty} \frac{ydy}{\sqrt{y^2+g^2}}  F(J_e^-(y))s^{J_e^-(y)} 
\end{eqnarray}
where, $J_{\pm}^+=-1\pm iy$.\\
 The analysis that follows now will actually mimic that done in  the previous subsection for zero magnon. But anyway we give the details step by step. What we do is to convert the above integral effectively into a complex contour integral as shown in the following figure. This is actually a \enquote{Wick rotation}   which we explain below. For further analysis we refer to the following figure.
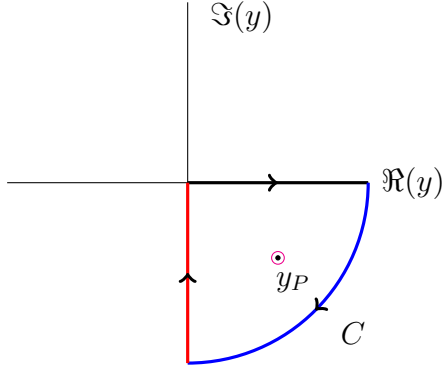
\begin{figure}[h]
	\centering
	\begin{tikzpicture}[scale=0.4]
	\draw[-] (-6,0)--(6,0);
	\draw[-] (0,-6)--(0,6);
	\draw (7.5,0) node {$\Re(y)$};
	\draw (1.8,5.5) node {$\Im(y)$};
	\draw (5.5,-5) node {$C$};
	\draw
	[
	very thick,
	decoration={markings, mark=at position 0.5 with {\arrow[black,line width=0.5mm]{>}}},
	postaction={decorate}
	]
	(0,0)--(6,0);
\draw
[
very thick,
color=blue,
decoration={markings, mark=at position 0.5 with {\arrow[black,line width=0.5mm]{>}}},
postaction={decorate}
]
(6,0) arc[start angle=0, end angle=-90, radius=6cm];
\draw
[
very thick,
color=red,
decoration={markings, mark=at position 0.50 with {\arrow[black,line width=0.5mm]{>}}},
postaction={decorate}
]
(0,-6)--(0,0);
\filldraw  (3,-2.5)circle[radius=2pt];
\draw[color=magenta] (3,-2.5)circle[radius=6pt];
\draw (3.5,-3.3) node {$y_P$};
\end{tikzpicture}
\caption{Contour Prescription for (A.14)}
\end{figure}\\\noindent
Referring to the above figure we can write our original integral as,
\begin{align} 
&\int_{0}^{\infty} \frac{ydy}{\sqrt{y^2+g^2}}  F(J_e^-(y))s^{J_e^-(y)}\nonumber\\
=& \int_{0}^{-i \infty} \frac{ydy}{\sqrt{y^2+g^2}}  F(J_e^-(y))s^{J_e^-(y)}-2\p i\sum_{y_P}\text{Res.}\left[ \frac{y}{\sqrt{y^2+g^2}}  F(J_e^-(y))s^{J_e^-(y)}\right]_{y=y_P}
\end{align} 
where,  $\lbrace y_P\rbrace$ are the  poles of  $F(J_e^-(y))$ in $y$.
Note that we have closed the contour in the lower half plane to ensure that the integral over $C$, which is a semi-circular arc of infinite radius, vanish. Also note that the residue sum comes with an overall negative sign because we have closed the contour in the clockwise sense. Now it is very clear from the above representation that only those poles which lie in the lower half-plane , as shown in the figure, i.e, the poles with negative imaginary parts can contribute to the residue sum i.e, the poles that contribute have the generic structure,
\be 
y_P=\Re(y_P)-i\bar{\Im}(y_P),\qquad \bar{\Im}(y_P)>0
\ee 
 And since each pole contributes  a factor of the form towards the residue,
$$ s^{-1-i y_P}$$
 it is immediately clear that these contributions, if any, have the form,
 \be 
 s^{-1-\bar{\Im}(y_P)+i\Re(y_P)}
 \ee 
  Clearly these contributions are exponentially suppressed compared to the line integral in the limit $s\rightarrow\infty$ i.e, the Regge limit. Hence forth while writing we will not write these pole contributions, if any, explicitly and \textit{any equality will be understood modulo  contributions coming from these poles}. Now we introduce $y=-iy_E$ (this is the \enquote{Wick rotation}\footnote{after Wick rotation $J_e^{\pm}=-1\pm y_E$} we mentioned above) and finally obtain,
\be 
I_1=-2\text{Re}\int_{0}^{\infty} \frac{y_Edy_E}{\sqrt{-y_E^2+g^2}}  F(J_e^-(y_E))s^{J_e^-(y_E)} 
\ee 
Now, let us look at the wick rotated part,
\be
I_1 = -2 \textrm{Re}\int_{0}^{g} \frac{y_Edy_E}{\sqrt{-y_E^2+g^2}}  F(J_e^-(y_E))s^{J_e^-(y_E)} -2 \textrm{Re}\int_{g}^{\infty} \frac{y_Edy_E}{\sqrt{-y_E^2+g^2}} F(J_e^-(y_E))s^{J_e^-(y_E)}
\ee
with, 

\begin{align}
F(J_e^-(y_E))& = \frac{
	16 c^4 \sqrt{g^2-y_E^2}  \sinh \left(\pi \sqrt{g^2-y_E^2}\right)}{\pi^2  g^2\sin(\pi  (y_E-1))}
\frac{
	 \Gamma \left(\frac{1-y_E-i \sqrt{g^2-y_E^2}}{2}\right) \Gamma \left(\frac{1-y_E+i \sqrt{g^2-y_E^2}}{2}\right)
}
{
	 \Gamma \left(\frac{-y_E-i \sqrt{g^2-y_E^2}}{2}\right)  \Gamma \left(\frac{-y_E+i \sqrt{g^2-y_E^2}}{2}\right) 
}\nonumber\\
&\hspace{2 cm}\times \G\left(\frac{3+y_E-t+i\sqrt{g^2-y_E^2}}{2}\right)\G\left(\frac{3+y_E-t-i\sqrt{g^2-y_E^2}}{2}\right)
\end{align}
To proceed further, we use a crucial observation about the \enquote{physical spectrum of $t$}. The vital information is that the physical spectrum for $t$ consists of \textit{real values only}. And henceforth we will base our analysis on the physical spectrum of $t$. With this piece of information we observe that the collections of Gamma functions come in the combination,
\be 
\G(p+iq)\G(p-iq),\qquad p,q\in\mathbb{R}
\ee 
with suitable values for $p,q$ (there are precisely three such combinations in the expression (A.28)). Since we have (this can be proved for instance using the Euler integral representation of Gamma function)
\be 
\G(z^*)=\G(z)^*
\ee 
so that 
\be 
\G(p+iq)\G(p-iq)=|\G(p+iq)|^2\,\in\mathbb{R}
\ee  \\
Hence, $F(J_e^-(y_E))$ is real over the entire interval $y\in [0,\infty)$. However the factor 
\be 
\frac{y_E}{\sqrt{-y_E^2+g^2}}
\ee  is purely real for $y_E\in [0,g]$ but is purely imaginary for $y_E\in [g,\infty)$. Thus the piece of integral in (A.6) over the interval $ [g,\infty)$ vanishes identically and we have therefore,
\be
I_1 = -2\int_{0}^{g} \frac{y_Edy_E}{\sqrt{-y_E^2+g^2}} F(J_e^-(y_E))s^{J_e^-(y_E)} 
\ee
On the other hand, now consider the integral
\be
I_2=-\int_{-g}^{g} d\n F(J_e^-)s^{J_e^-}
\ee
Under the transformation $-\nu^2+g^2= \tilde{y}^2$, we have $J_{-}^e=-1-\tilde{y}$ and ,

\begin{eqnarray}
I_2&=& -2\int_{0}^{g} d\n F(J_e^-)s^{J_e^-} \nonumber\\
&=& -2\int_{0}^{g}  \frac{\tilde{y} d\tilde{y}}{\sqrt{g^2-\tilde{y}^2}}F(J_e^-(\tilde{y}))s^{J_e^-(\tilde{y})} 
\end{eqnarray}
Now note that $F(J_-^+(\tilde{y}))$ above is same as $F(J_-^+(y_E))$ in (A.28) with the replacement $y_E\rightarrow \tilde{y}$. Thus we have the relation, 
\begin{eqnarray}
I_1 =  I_2 
\end{eqnarray}
Equipped with this we have the following identities,
\begin{eqnarray}
\int_{-\infty}^{-g} d\n F(J_e^+)\left(\frac{s}{4}\right)^{J_e^+} + \int_{g}^{\infty} d\n F(J_e^-)s^{J_e^-}  = -\int_{-g}^{fg} d\n F(J_e^-)s^{J_e^-}\,.
\end{eqnarray}
\begin{eqnarray}
\int_{-\infty}^{-g} d\n F(J_e^-)\left(\frac{s}{4}\right)^{J_e^-} + \int_{g}^{\infty} d\n F(J_e^+)s^{J_e^+}  = -\int_{-g}^{g} d\n F(J_-)s^{J_e^-} \,.
\end{eqnarray}
Finally,
we add them together to arrive at , 
\be
{\cal{M}}_{(1)}^{+}(s,t) =\int_{-g}^{g} d\n \left( F(J_e^+)s^{J_e^+}- F(J_e^-)s^{J_e^-} \right) \,.
\ee
\section{Details of various integrals }\label{intweak}
We note that in zero magnon and one magnon weak coupling case we finally are left with evaluation of the integrals of the form
\be 
\mathcal{I}_{n}(P)=\int_{-1}^{1}dx e^{P x} \sqrt{1-x^2} x^n\,,\qquad n\in\mathbb{Z}
\ee 
Now we can generate all such integrals from the basic integral by repeated applications of derivative (for non-negative $n$) or anti derivative (for negative $n$) with respect to $L$ of the the following basic integral,
\be 
\mathcal{I}_{0}(P)=\int_{-1}^{1}dx e^{P x} \sqrt{1-x^2}=\frac{\pi  I_1(P)}{P}
\ee 
where $I_\m(L)$ is Modified Bessel function of first kind.\\
\paragraph{} For non-negative $n$, we have  the following differential relation,
\be \label{npos}
\mathcal{I}_{n}(P)=\frac{d^n}{dP^n}\mathcal{I}_{0}(P),\qquad n\geq0
\ee
with $n=0$ corresponds to no differentiation.\\
For example, 
\be 
\mi_1(P)=\frac{d}{dP}\mi_0(P)=\p\frac{I_2(P)}{P}
\ee 
\paragraph{} On the other hand we note that for $n<0$ the integrand is singular at $x=0$. So in this case the integral as such does not exist. However the integral can still be given meaning in the sense of Cauchy Principal value. Thus we have the following integral under consideration,
\be 
\tilde{\mathcal{I}}_{n}(P)=\text{P.V.}\int_{-1}^{1}dx e^{P x} \sqrt{1-x^2} x^n=\lim_{\d\rightarrow0}\left[\int_{-1}^{\d}+\int_{\d}^{1}\right]dx e^{P x} \sqrt{1-x^2}x^n,\quad n\in\mathbb{Z}^-
\ee 
We can get this integral from $\mathcal{I}_0(L)$ by repeated anti derivative operation i.e, repeated indefinite integral w.r.t $L$. Thus if we define,
\be 
\ml=\int dP
\ee 
then,
\be \label{nneg}
\tilde{\mi}_n(P)=\ml^n\mi_0(P)=\int^PdP_n\int^{P_n}dP_{n-1}\ldots\int^{P_2}dL_1\,\mi_0(P_1)
\ee 
For example ,
\be 
\tilde{\mi}_{-1}(P)=\int^{P}dP_1\mi_0(P_1)=\frac{\p}{2} P \, _1F_2\left(\frac{1}{2};\frac{3}{2},2;\frac{P^2}{4}\right)
\ee 
This can be expressed in terms of modified Bessel functions and modified Struve functions as following,
\be 
\tilde{\mi}_{-1}(P)=\frac{\p}{2} (P (\pi  \pmb{L}_1(P)+2) I_0(L)-(\pi  P \pmb{L}_0(P)+2) I_1(P))
\ee 
where, $I_{\m}(z)$ is modified Bessel function of first kind and $\pmb{L}_\n(z)$ is modified Struve function. In general $\tilde\mi_{-n}(P), n>0$ can be expressed in terms of  Bessel functions and  Struve functions.
\section{Details of 2-Magnon Analysis}\label{2magdetails}
In this appendix we give the details of , 
\be\label{A1}
A_1=\frac{\zeta_2(\D_i,t)}{2\pi i}\left(\int_{-\infty}^{-\x^4}+\int_{\x^4}^\infty\right) d\n \n\sinh\pi \n\oint dJ \frac{\p}{\sin \pi J}\mathfrak{M}(J,\n)\left(\frac{s}{4}\right)^J \left(\frac{E^{(2)}_{2+i\n,J}}{1-\xi^4 E^{(2)}_{2+i\n,J}}\right)_{|\xi|<1,|\n|>1}
\ee
with $$\zeta_2(\D_i,t)=\frac{1}{\G(2-\frac{t}{2})^2}.$$
The general solution in this regime is given by \eqref{pert0} which for convenience and generality, we can write,
\be
J_n^a(\pm\n)=\pm i\n-a-4n+\sum_{k\geq1} \xi^{2k}\g_n^a(\pm\n)\,,
\ee
for $a=2,4$. Further, we denote the Jacobian of transformation as,
\be
\Theta_n^a(\n)=\bigg|\frac{\pd J_n^a(\n)}{\pd\g_n^a(\n)}\bigg|\,.
\ee
Now we have for the integral $A_1$,after doing the $J$ intgral,
\be 
A_1=\zeta_2(\D_i,t)\sum_{m=0}^{\infty}\left(\int_{-\infty}^{-\x^4}+\int_{\x^4}^\infty\right) d\n \left[\mathfrak{F}(J_m^2(\pm\n),\n)+\mathfrak{F}(J_m^4(\pm\n),\n)\right]
\ee
where, 
\begin{align} 
\begin{split}
\mathfrak{F}(J_n^a(\pm\n),\n)=\n\sinh\pi \n\frac{\Theta_n^a(\pm\n)}{\sin\pi J_n^a(\pm\n)}\mathfrak{M}(J_n^a(\pm\n),\n)\left(\frac{s}{4}\right)^{J_n^a(\pm\n)}\left(\frac{E^{(2)}_{2+i\n,J}}{1-\xi^4 E^{(2)}_{2+i\n,J}}\right)_{J=J_n^a(\pm\n)}\,,
\end{split}
\end{align}
Putting all the expressions one obtains the following perturbative expressions,
\begin{align} 
\begin{split}
\mathfrak{F}(J_m^2(\pm\n),\n)&=- \frac{L^{-4 m\pm i \nu -2}((2 m)!)^2  \Gamma \left(2 m-\frac{t}{2}+2\right)\Gamma (\pm2 i \nu -4 m)\Gamma \left(2 m\mp i \nu -\frac{t}{2}+2\right) }{128 \pi ^{9}  (4 m)!\Gamma (\pm i \nu -2 m)^2 }+O(\x^4),
\end{split}
\end{align}
\begin{align} 
\begin{split}
\mathfrak{F}(J_m^4(\pm\n),\n)&=\frac{L^{-4 m\pm i \nu -4}\Gamma (2 m+2)^2   \Gamma \left(2 m-\frac{t}{2}+3\right)\Gamma (2\pm i \nu-4 m -2)\Gamma \left(2 m\mp i \nu -\frac{t}{2}+3\right) }{128 \pi ^{9} \Gamma (4 m+3) \Gamma (\pm i \nu-2 m -1)^2}+O(\x^4).
\end{split}
\end{align}
with $L=(s/4)$. We modify our integral as follows,
\be
\begin{split}
	\frac{A_1}{\zeta_2(\D_i,t)}&=\int_{-\infty}^\infty\ d\n  \sum_{m=0}^\infty \sum_{a=2,4}\mathfrak{F}(J_m^a(\n),\n)
	-\int_{-\x^4}^{\x^4}d\n  \sum_{m=0}^\infty\sum_{a=2,4}\mathfrak{F}(J_m^a(\n),\n)~+~~(\n\to-\n)\\
	&=I_1 -I_2
\end{split}
\ee
with,
\begin{align}
\begin{split} 
I_1&=\int_{-\infty}^\infty\ d\n\sum_{m=0}^\infty \sum_{a=2,4}\mathfrak{F}(J_m^a(\n),\n)~+~~(\n\to-\n)\\
I_2&=\int_{-\x^4}^{\x^4}d\n  \sum_{m=0}^\infty\sum_{a=2,4}\mathfrak{F}(J_m^a(\n),\n)~+~~(\n\to-\n)
\end{split}
\end{align} 
We evaluate the integrals for the leading Regge trajectory, namely $a=2,m=0$, above. Therefore we will focus upon,
\be 
\bar{I}_{1(2)}(s,t)=\int_{-\infty(-\x^4)}^{+\infty(+\x^4)}d\n~ \left[- \frac{L^{ i \nu -2}  \Gamma \left(-\frac{t}{2}+2\right)\Gamma (2 i \nu )\Gamma \left( -i \nu -\frac{t}{2}+2\right) }{128 \pi ^{9}  \Gamma ( i \nu )^2 }+O(\x^4)\right]+(\n\to-\n).
\ee 
\subsection*{Evaluating $\bar{I}_1(s,t)$}We start with,
\begin{align}
\begin{split} 
\bar{I}_1(s,t)=\int_{-\infty}^\infty\ d\n  \mathfrak{F}(J_0^2(\n),\n)=\frac{1}{64\p^{9}} \left[
- \Gamma \left(-\frac{t}{2}+2\right)L^{-2}\int_{-\infty}^{\infty}d\n L^{i \nu}\mi_0^2(\n,t)
\right]+O(\x^4)
\end{split}
\end{align}
with $L= (s/4)$ and,
\be
\mathcal{I}_{0}^2(\n;t)= \frac{  \Gamma (2 i \nu )  \Gamma \left(-i \nu -\frac{t}{2}+2\right)}{  \Gamma (i \nu )^2}.
\ee 
Note that we have used In order to do each of the above integrals we will resort to contour integral. First we will consider the following contour integral,
\be 
\oint_{\g}d\n L^{i\n}\, \mathcal{I}_{0}^2(\n;t)=\lim_{R\rightarrow\infty}\int_{-R}^{R}d\n L^{i\n}\, \mathcal{I}_{0}^2(\n;t)+\lim_{R\rightarrow\infty}\int_{C_R}d\n L^{i\n}\, \mathcal{I}_{0}^2(\n;t)
\ee 
where $C_R$ is a semi-circular arc centered at the origin, having a radius of $R$ and traversed in the counter-clockwise direction  . The arc lies in the upper half $\n$ plane, i.e, with $\Im(\n)>0$. Further in the limit that radius of the semicircular arc $C_R$ goes to infinity,
\be 
\lim_{R\rightarrow\infty}\int_{C_R}d\n L^{i\n}\, \mathcal{I}_{0}^2(\n;t)\rightarrow0
\ee
Thus we have ,
\be 
\int_{-\infty}^{\infty}d\n L^{i\n}\, \mathcal{I}_{0}^2(\n;t)=\oint_{\g}d\n L^{i\n}\, \mathcal{I}_{0}^2(\n;t)
\ee 
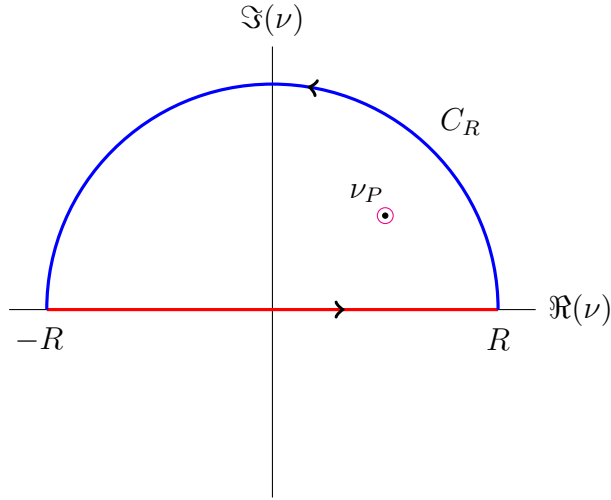
\begin{figure}[h]
	\centering
	\begin{tikzpicture}[scale=0.5]
	\draw[-] (-7,0)--(7,0);
	\draw[-] (0,-5)--(0,7);
	\draw (8.2,0) node {$\Re(\n)$};
	\draw (0,7.7) node {$\Im(\n)$};
	\draw (5,5) node {$C_R$};
	\draw (-6.2,-.8) node {$-R$};
	\draw (6,-.8) node {$R$};
	\draw
	[
	very thick,
	color=red,
	decoration={markings, mark=at position 0.66 with {\arrow[black,line width=0.5mm]{>}}},
	postaction={decorate}
	]
	(-6,0)--(6,0);
	\draw
	[
	very thick,
	color=blue,
	decoration={markings, mark=at position 0.45 with {\arrow[black,line width=0.5mm]{>}}},
	postaction={decorate}
	]
	(6,0) arc[start angle=0, end angle=180, radius=6cm];
	\filldraw  (3,2.5)circle[radius=2pt];
	\draw[color=magenta] (3,2.5)circle[radius=6pt];
	\draw (2.5,3.1) node {$\n_P$};
	\end{tikzpicture}
	\caption{Contour Integral for (C.2)}
\end{figure}\\\noindent
And therefore we focus our attention towards doing the contour integral for which we will do pole analysis for each integrand $\mathcal{I}_m^a$ in order to take advantage of the Residue theorem,
\be 
\oint_{\g}d\n L^{i\n}\, \mathcal{I}_{0}^2(\n;t)=2\p i\sum_{\n_P} \text{Res.}[L^{i\n}\mathcal{I}_{0}^2(\n;t)]_{\n=\n_P}
\ee 
where $\n_p$ are the $\n$ poles enclosed within the contour $\g$.
Now clearly the poles that can contribute to this integral must lie in the upper half plane as shown in the figure above i.e, such a $\n_P$, if any , must have positive imaginary part in order to contribute to (C.7). Hence such a pole has to have the generic form,
\be 
\n_P=\Re(\n_P)+i\Im(\n_P),\qquad \Im(\n_P)>0
\ee 
For completeness let us locate these poles of $\mi_0^2(\n)$. There are two kinds of poles of the $\n-$integrand. These are as following, 
\begin{enumerate}
\item {\bf $t-$ independent poles:} These poles are given by ,
\be \label{tind}
\n_P^{(1)}(q)=i\frac{2q+1}{2},~~~~q\in\mathbb{Z}/\mathbb{Z}^-.
\ee 	
The residue at these poles are given by, 
\be 
\sum_{\n_P^{(1)}(q)}\text{Res.}[L^{i\n}\mathcal{I}_{0}^2(\n;t)]_{\n=\n_P^{(1)}(q)}=\sum_{q} \frac{i \Gamma \left(q+\frac{3}{2}\right)^2 \Gamma \left(q-\frac{t}{2}+\frac{5}{2}\right)}{2 \pi ^2 \Gamma (2q+2)}L^{-\frac{2q+1}{2}}
\ee 
\item {\bf $t-$dependent poles:} These pole locations are given by, 
\be \label{tdep}
\n_P^{(2)}(q;t)=i\left(\frac{t}{2}-2-q\right),~~~~q\in\mathbb{Z}/\mathbb{Z}^-.
\ee 
Now these poles will contribute to the contour integral according to the values of $t$.  By values of $t$, we are ultimately interested in \enquote{physically admissible} values of $t$. These are obtained from the poles of $\Gamma \left(-\frac{t}{2}+2\right)$. These poles are at, 
\be \label{tpole}
t_j=2(j+2),~~~~j\in\mathbb{Z}/\mathbb{Z}^-.
\ee 
Putting these into \eqref{tdep} we have, 
\be 
\n_P^{(2)}(q;t_j)=i(j-q)
\ee 
Now we clearly see that only when $j> q$ then these poles will contribute. Thus we have, 
\be 
\sum_{\n_P^{(1)}(q)}\text{Res.}[L^{i\n}\mathcal{I}_{0}^2(\n;t_j)]_{\n=\n_P^{(2)}(q;t_j)}=\sum_{q<j}\frac{i (-1)^q \Gamma (2 q-2 j)}{q! \Gamma (q-j)^2}L^{-(j-q)}
\ee 
\end{enumerate}
Further putting $q=j-\ell,\ell>0$ we readily see that, each summand of this residue sum vanishes identically. 
Therefore, collecting everything together, we are left with, 
\be 
\bar{I}_1(s,t)=\frac{1}{64\p^{10}} \left[
\Gamma \left(-\frac{t}{2}+2\right)L^{-2}\sum_{q} \frac{ \Gamma \left(q+\frac{3}{2}\right)^2 \Gamma \left(q-\frac{t}{2}+\frac{5}{2}\right)}{ \Gamma (2q+2)}L^{-\frac{2q+1}{2}}\right]+O(\x^4)
\ee 
Thus in the Regge limit,  $s\to\infty,$ with $t$ fixed, we can write the asymptotic equivalence relation, 
\be \label{I1}
\bar{I}_1(s,t)\sim \frac{1}{8\p^{9}}\Gamma \left(\frac{4-t}{2}\right)\G\left(\frac{5-t}{2}\right) s^{-\frac{5}{2}}+O(\x^4),~~~s\to\infty
\ee 
Note that while we have done this analysis with explicitly the $O(\xi^0)$ expression the same conclusion will hold true for higher orders because in higher order basically we will encounter higher order Polygamma functions with the argument, however, unchanged. Thus the pole locations in complex $\n$ plane won't be altered.
\subsection*{Evaluation of $\bar{I}_2(s,t)$}
Next we move on to evaluating, 
\be 
\bar{I}_2(s,t) =-\frac{1}{64\p^{9}}
\Gamma \left(-\frac{t}{2}+2\right)L^{-2}\int_{-\x^4}^{\x^4}d\n \ e^{i\l \nu}\frac{\Gamma (2 i \nu )\Gamma \left(- i \nu -\frac{t}{2}+2\right) }{  \Gamma ( i \nu )^2 }
\ee 
where we have defined, $\l=\log L$. Note that as $L\rightarrow \infty$ so does $\l$ i.e, $\l\rightarrow \infty$ however at a much slower rate. We can do this integral by parts repeatedly.

So as far as the integral is concerned, we are dealing with a \enquote{Stationary Phase} type of configuration. The integral can be written as ,
\be 
\int_{a}^{b}d\n e^{i\l \phi(\n)} \L(\n)
\ee 
with $\l\rightarrow \infty, a=-\x^4, b=\x^4, \phi(\n)=\n$ and
\be 
\L(\n)=\mi_m^2(\n;t)=\frac{  \Gamma (2 i \nu )  \Gamma \left(-i \nu -\frac{t}{2}+2\right)}{  \Gamma (i \nu )^2}
\ee 
In order to tackle this kind of problem one normally looks for stationary points i.e, $\n$ values in $[a,b]$ such that $\phi'(\n)=0$. However in our case we see that  $\phi'(\n)=1\neq 0$ identically in $[-1,1]$. So what we will resort to is integration by parts.By integration by parts one can obtain the general expression,
\be 
\int_{-\x^4}^{\x^4}d\n e^{i\l\n}\psi(\n)=\frac{1}{i\l}\left[e^{i\l\x^4}\sum_{n=0}^{N-1}\frac{\L_n(\x^4)}{(i\l)^n}-e^{-i\l\x^4}\sum_{n=0}^{N-1}\frac{\L_n(\x^4)}{(i\l)^n}\right]+\frac{1}{(i\l)^N}\int_{-\x^4}^{\x^4}e^{i\l\n}\L_N(\n) d\n
\ee 
where we have $\psi_0=\psi$ and, 
\be 
\L_{n+1}(\n)=-\frac{d}{d\n}\L_n(\n),~~~~n=0,1,2,\dots.
\ee 
Now at this point one can go upto $N$ as desired and then if one expands the result around $\x=0$ then one will generate a perturbation series in $\x$.  For a specific illustration let us choose $N=3$. However we would like to point to certain universal behaviour of the  perturbative expansion. Therefore we start from,
\be 
\int_{-\x^4}^{\x^4}d\n e^{i\l\n}\L(\n)=\frac{1}{i\l}\left[e^{i\l\x^4}\sum_{n=0}^{2}\frac{\L_n(\x^4)}{(i\l)^n}-e^{-i\l\x^4}\sum_{n=0}^{2}\frac{\L_n(\x^4)}{(i\l)^n}\right]+\frac{1}{(i\l)^3}\int_{-\x^4}^{\x^4}e^{i\l\n}\L_3(\n) d\n.
\ee 
We expand the parethesized term above in $\x$ upto order $\x^{12}$ to obtain, 
\begin{align}
\begin{split}
&\frac{1}{i\l}\left[e^{i\l\x^4}\sum_{n=0}^{2}\frac{\L_n(\x^4)}{(i\l)^n}-e^{-i\l\x^4}\sum_{n=0}^{2}\frac{\L_n(\x^4)}{(i\l)^n}\right]\\
=&-\frac{\xi ^4 \Gamma \left(2-\frac{t}{2}\right)^2 }{64 \pi ^9 \lambda ^3}\left[3 \psi ^{(0)}\left(2-\frac{t}{2}\right)^2+3 \psi ^{(1)}\left(2-\frac{t}{2}\right)+\pi ^2\right]+\frac{\xi ^{12}\Gamma \left(2-\frac{t}{2}\right)^2}{192\p^9}\left[\l-6 \psi ^{(0)}\left(2-\frac{t}{2}\right)+O(\l^{-1})\right].\\
\end{split}
\end{align}

Now at this point we would like to mention some general characteristics. In Regge limit the $\x^{12}$ term has the leading contribution proportional to $\log L$. If we went to to higher orders in $\x$ we would have gotten higher powers of $\log L$, for example order $\x^{20}$ term will give a contribution proportional to $\log^2L$. Further if one choses greater values of $N$ then one will get $\x^4$ term suppressed more and more by negative powers of $\log s$. So we can for practical purpose dispense with such suppressed terms. So we can write to the leading contribution to order $\x^{12}$ in Regge limit $L\to\infty$,
\be \label{I2}
\bar{I}_2(s,t)\sim \xi ^{12}\frac{\Gamma \left(2-\frac{t}{2}\right)^2}{192\p^9L^2}\left[\log L-6 \psi ^{(0)}\left(2-\frac{t}{2}\right)\right]+O(\x^{20}),~~~L\to\infty.
\ee
Now if we assume further that $L$ is the largest scale in the theory so that upto a given order in $\x$ expansion  the dominant contribution in Regge limit comes from the maximal power of $L$ then $\bar{I}_1(s,t)$ will always be exponentially suppressed compared to $\bar{I}_2(s,t)$. Thus for the  leading Regge trajectory considered in section \ref{inteva} we can write,
\be 
A_1\sim -\frac{\xi ^{12}}{192\p^9L^2}\left[\log L-6 \psi ^{(0)}\left(2-\frac{t}{2}\right)\right]+O(\x^{20}).
\ee 

\providecommand{\href}[2]{#2}\begingroup\raggedright\endgroup

\end{document}